\documentclass[preprint,authoryear,sort&compress,12pt]{elsarticle}  
\usepackage[top=1.25in, bottom=1.25in, left=0.75in, right=0.75in]{geometry}
\usepackage{graphicx}
\usepackage{amssymb}
\usepackage{amsmath}
\usepackage{lineno}
\usepackage{subfig}

\usepackage{color}
\usepackage{soul}
\usepackage{array}
\usepackage{hyperref}
\usepackage[colorinlistoftodos]{todonotes}
\usepackage{makecell}

\newcommand{\xmb}[1]{\ensuremath{\mathbf{#1}}}
\newcommand{\xmbs}[1]{\ensuremath{\boldsymbol{#1}}}

\journal{}

\begin{document}

\begin{frontmatter}

\title{Sediment Micromechanics in Sheet Flows Induced by Asymmetric Waves: A CFD--DEM Study}

 \author{Rui Sun} \ead{sunrui@vt.edu}
 \author{Heng Xiao\corref{corxh}} \ead{hengxiao@vt.edu}
 \address{Department of Aerospace and Ocean Engineering, Virginia Tech, Blacksburg, VA 24060, United
 States}

 \cortext[corxh]{Corresponding author. Tel: +1 540 231 0926}

\begin{abstract}
  Understanding the sediment transport in oscillatory flows is essential to the investigation of the
  overall sediment budget for coastal regions. This overall budget is crucial for the prediction of
  the morphological change of the coastline in engineering applications. Since the sediment
  transport in oscillatory flows is dense particle-laden flow, appropriate modeling of the particle
  interaction is critical. Although traditional two-fluid approaches have been applied to the study
  of sediment transport in oscillatory flows, the approaches do not resolve the interaction of the
  particles. Particle-resolved modeling of sediment transport in oscillatory flows and the study of
  micromechanics of sediment particles (e.g., packing and contact force) are still lacking. In this
  work, a parallel CFD--DEM solver \textit{SediFoam} that can resolve the inter-particle collision
  is applied to study the granular micromechanics of sediment particles in oscillatory flows. The
  results obtained from the CFD--DEM solver are validated by using the experimental data of coarse
  and medium sands.  The comparison with experimental results suggest that the flow velocity, the
  sediment flux and the net sediment transport rate predicted by \textit{SediFoam} are satisfactory.
  Moreover, the micromechanic quantities of the sediment bed are presented in detail, including the
  Voronoi concentration, the coordination number, and the particle interaction force. It is
  demonstrated that variation of these micromechanic quantities at different phases in the
  oscillatory cycle is significant, which is due to different responses of the sediment bed. To
  investigate the structural properties of the sediment bed, the correlation of the Voronoi volume
  fraction and coordination number are compared to the results from the fluidized bed simulations.
  The consistency in the comparison indicates the structural micromechanics of sediment transport
  and fluidized bed are similar despite the differences in flow patterns. From the prediction of the
  CFD--DEM model, we observed that the coordination number in rapid sheet flow layer is larger than
  one, which indicates that a typical particle in the sediment layer is in contact with more than
  one particles, and thus the binary collision model commonly used in two-fluid approaches may
  underestimate the contact between the particles.
\end{abstract}


\end{frontmatter}


\section{Introduction}
\label{sec:intro}


Oscillatory sheet-flow sediment transport is important in coastal and geotechnical engineering
because it carries significant amount of sand and has a considerable impact on the overall sediment
budget~\citep{malarkey09mo}. Since the inter-particle collision plays an important role in
oscillatory sheet-flow sediment transport, understanding the mechanics at the particle scale can
provide physical insights of this problem. The microscopic information can describe the properties
of the sediment bed, such as packing, mixing, and permeability~\citep{yi11cn}. In addition, the
structural strength of the seabed, without which the sediment particles will behave as
fluid~\citep{scholtes15mw}, can be also evaluated by using micromechanic variables. There is also an
increasing interest to seek for the constitutive laws for dense granular flows in the macroscopic
modeling of sediment transport problems based on the microscopic information, including the
understanding of the anisotropic nature of the soil, the response of the sediment bed under stress,
and the relationship between particle-scale response and material response~\citep{o11pdem}.

\subsection{Sediment Transport in Oscillatory Sheet-Flow}

The wave orbital motion is dominant in cross-shore sediment transport~\citep{ribberink98bt}. The
wave-induced oscillatory flow may lead to onshore or offshore sediment transport, and thus is
important in the prediction of beach-profile changes. Although some researchers performed
experimental studies of the sediment transport in oscillatory flows, the measurement of the
concentration and velocity of sediment particles are still considered very
difficult~\citep{flores98mb,od04co,od04ft}. For example, ~\cite{od04ft} used both conductivity
concentration measurement (CCM) probes and suction samplers to measure the sediment concentration.
Conductivity concentration probes are used to measure the sediment concentration in the region of
high concentration in the sheet flow layer, whereas suction samplers are used in the low
concentration region in the suspension layer. The samplers are located at approximately one
centimeter above the sediment bed to avoid the possible interaction of suction sampler and sediment
bed. Therefore, the regions with medium concentration cannot be measured and there is a gap above
the sediment bed in the experimental measurement data. Because of the difficulties in the
experimental measurements, numerical simulations have been a cost-effective approach to study the
sediment transport in oscillatory flows.

A traditional numerical simulation approach is the two-fluid
model~\citep{hanes85gf,jenkins98cs,hsu04two,malarkey09mo,cheng14ts}, in which both the fluid and
particle phases are described as inter-penetrating continuum. This approach does not capture
the motion of individual sediment particle but solves the concentration field of the particles. To
model the inter-particle collision, constitutive relations based on binary collision
assumptions are used~\citep{cheng14ts}. The two-fluid model is satisfactory in capturing the
sediment concentration and sediment flux compared to the experimental
results~\citep{malarkey09mo,cheng14ts}. However, since two-fluid models rely on constitutive
relations of the particle phase based on binary collisions, it is only valid in relatively dilute
flows. In dense flows, the particles have endured contact, and thus the binary collisions may not be
applicable~\citep{o11pdem,hou12mm}.

CFD--DEM (Computational Fluid Dynamics--Discrete Element Method) is another numerical approach that
has been applied to study sediment transport in sheet flows. In contrast to the two-fluid model,
CFD--DEM models the motion of sediment particles explicitly, and accounts for the collision
of sediment particles. However, due to the limitation of the computational costs, the study of
sediment transport in oscillatory flows using CFD--DEM was simplified in the early
studies~\citep{drake01dp,calantoni04ms}. The oscillatory flow was modeled as two-dimensional layers
and only a few thousand particles were used. In addition, the early studies did not resolve the
instantaneous turbulent flow and only investigated the mechanics of very coarse sediment particles
in oscillatory sheet-flow. With the growth of available computational resources, this approach has
gained more popularity in solving sediment transport problems. In the recent
works~\citep{schmeeckle14ns,sun16cfd}, the motion of medium or coarse sand particles in turbulent
flows is captured.

\subsection{Granular Micromechanics}

The granular micromechanics of particle-laden flow are critical in the physical understanding of
granular microscopic structures, especially for chemical and pharmaceutical engineering
problems~\citep{kuang11mm,hou12mm}. The micromechanic variables, including Voronoi volume fraction,
coordination number, contact force and others quantities, are obtained based on the packing of the
particles.  Since such micromechanic variables are much easier to produce from DEM than measuring
from experiment, DEM has been extensively used to investigate the granular micromechanics in the
past decade~\citep{yi11cn}.

DEM has been widely applied to investigate the granular micromechanics in engineering applications,
for example, hopper, rotating drum, and sand piles~\citep{langston95dem,yang23ma,zhao13coup}.
Micromechanic variables are investigated to understand the heat conduction, particle agglomeration
and structural stability. CFD--DEM simulations of pneumatic convey and fluidized bed problems are
also performed to study the variation of micromechanic variables in different flow
regimes~\citep{kuang11mm, hou12mm}. Moreover, the constitutive models are proposed according to the
micromechanics of the granular materials~\citep{sun11jfm}, which contributes to macroscopic
modeling. Previous studies of granular micromechanics have demonstrated that the CFD--DEM approach
has good potential for the granular micromechanics in geotechnical engineering, for example, the
modeling of arbitrary-shaped particles~\citep{o11pdem}. However, the study in granular
micromechanics of sediment transport is still lacking.

Although recent studies prove that CFD--DEM is capable of predicting sediment transport in
unidirectional turbulent flow~\citep{kidanemariam14dn,schmeeckle14ns,sun16cfd}, the capability of
this approach in predicting the motion of coarse and medium sand in oscillatory flow is still not
demonstrated. The present work aims at (1) demonstrating that CFD--DEM is capable of predicting the
instantaneous motion of sediment particles in oscillatory sheet-flow, and (2) investigating the
granular micromechanics in sediment transport process. Coarse and medium sands are used in the
present numerical simulations. The instantaneous turbulent flow field within the boundary layer is
resolved.

The rest of the paper is organized as follows. The methodology of the present model is introduced in
Section~\ref{sec:cfddem}, including the mathematical formulation of fluid equations, the particle
motion equations, the fluid--particle interactions, and diffusion-based averaging procedure. The
implementation of the code and the numerical methods used in the simulations are detailed in
Section~\ref{sec:num-method}. In Section~\ref{sec:simulations}, the results obtained in the
simulations are discussed. Finally, Section~\ref{sec:conclude} concludes the paper.

\section{Methodology}
\label{sec:cfddem}

\subsection{Mathematical Model of Particle Motion}
\label{sec:dem-particle}

In \textit{SediFoam}, the translational and rotational motions of each particle are calculated
based on Newton's second law as the following equations~\citep{cundall79,ball97si}: 
\begin{subequations}
 \label{eq:newton}
 \begin{align}
  m \frac{d\xmb{u}}{dt} &
  = \xmb{f}^{col} + \xmb{f}^{fp} + m \xmb{g} \label{eq:newton-v}, \\
  I \frac{d\xmbs{\Psi}}{dt} &
  = \xmb{T}^{col} + \xmb{T}^{fp} \label{eq:newton-w},
 \end{align}
\end{subequations}
where \( \xmb{u} \) is the velocity of the particle; $t$ is time; $m$ is particle mass;
\(\xmb{f}^{col} \) represents the contact forces due to particle--particle or particle--wall
collisions; \(\xmb{f}^{fp}\) denotes fluid--particle interaction forces; \(\xmb{g}\) is body
force. \(I\) and \(\xmbs{\Psi}\) are angular moment of inertia and angular velocity of the particle;
\(\xmb{T}^{col}\) and \(\xmb{T}^{fp}\) are the torques due to contact forces and fluid--particle
interactions, respectively.  To compute the collision forces and torques, the particles are modeled
as soft spheres with inter-particle contact represented by an elastic spring and a viscous dashpot.
Further details can be found in the study by~\cite{tsuji93}.

\subsection{Locally-Averaged Navier--Stokes Equations for Fluids}
\label{sec:lans}

The fluid flow is described by the locally-averaged incompressible Navier--Stokes equations.
Assuming constant fluid density \(\rho_f\), the governing equations for the fluid
are~\citep{anderson67,kafui02}:
\begin{subequations}
 \label{eq:NS}
 \begin{align}
  \nabla \cdot \left(\varepsilon_s \xmb{U}_s + {\varepsilon_f \xmb{U}_f}\right) &
  = 0 , \label{eq:NS-cont} \\
  \frac{\partial \left(\varepsilon_f \xmb{U}_f \right)}{\partial t} + \nabla \cdot \left(\varepsilon_f \xmb{U}_f \xmb{U}_f\right) &
  = \frac{1}{\rho_f} \left( - \nabla p + \varepsilon_f \nabla \cdot \xmbs{\mathcal{R}} + \varepsilon_f \rho_f \xmb{g} + \xmb{F}^{fp}\right), \label{eq:NS-mom}
 \end{align}
\end{subequations}
where \(\varepsilon_s\) is the solid volume fraction, \( \varepsilon_f = 1 - \varepsilon_s \) is the
fluid volume fraction, and \( \xmb{U}_f \) is the fluid velocity. The terms on the right hand side
of the momentum equation are: pressure (\(p\)) gradient, divergence of the stress tensor \(
\xmbs{\mathcal{R}} \) (including viscous and Reynolds stresses), gravity, and fluid--particle
interactions forces, respectively.  In the present study, we used large-eddy simulation to resolve
the flow turbulence in the computational domain. The stress tensor is composed of both viscous and
Reynolds stresses: \( \xmbs{\mathcal{R}} = \mu\nabla \xmbs{U_f} + \rho_f \mathcal{T}\), where $\mu$
is the dynamic viscosity of the fluid flow and $\mathcal{T}$ is the Reynolds stress. The expression
of the Reynolds stress is:
\begin{equation}
  \mathcal{T} = \frac{2}{\rho_f}\mu_t \xmbs{S} - \frac{2}{3}k\xmbs{I},
  \label{eq:reynolds-stress}
\end{equation}
where $\mu_t$ is the dynamics eddy viscosity, $\xmbs{S} = (\nabla \xmbs{U_f} + (\nabla
\xmbs{U_f})^T)/2$, and $k$ is the turbulent kinetic energy.  It is noted that in the stress tensor
\( \xmbs{\mathcal{R}} \) term, the fluctuations of the fluid flow at the boundary of the particle
are not resolved. We applied the one-equation eddy viscosity model proposed by~\cite{yoshizawa85sd}
as the sub-grid scale (SGS) model. This model is selected because it is proven to be adequate to
simulate the turbulent flow in a channel~\citep{horiuti85les}, and it is widely used in large-eddy
simulations based on OpenFOAM. In addition, ~\cite{yoshizawa85sd} observed that the standard
Smagorinsky model would be recovered from this model if production equals dissipation in the SGS
kinetic energy equation. 

\subsection{Fluid--Particle Interactions}
\label{sec:fpi}
The fluid-particle interaction force \(\xmb{F}^{fp}\) consists of buoyancy \( \xmb{F}^{buoy} \),
drag \( \xmb{F}^{drag} \), and lift force \(\xmb{F}^{lift}\). The drag on an individual particle
$i$ is formulated as:
\begin{equation}
  \mathbf{f}^{drag}_i = \frac{V_{p,i}}{\varepsilon_{f, i} \varepsilon_{s, i}} \beta_i \left( \mathbf{u}_{p,i} -
  \mathbf{U}_{f, i} \right),
  \label{eqn:particleDrag}
\end{equation}
where subscripts $p$ and $f$ indicate particle and fluid quantities; \( V_{p, i} \) and \(
\mathbf{u}_{p, i} \) are the volume and the velocity of particle $i$, respectively; \(
\mathbf{U}_{f, i} \) is the fluid velocity interpolated to the center of particle $i$; \( \beta_{i}
\) is the drag correlation coefficient which accounts for the presence of other particles.  The
$\beta_i$ value for the drag force is based on~\cite{mfix93}:
\begin{equation}
\beta_i = \frac{3}{4}\frac{C_d}{V^2_r}\frac{\rho_f |\xmbs{u_{p,i}} - \xmbs{U_{f,i}}|}{d_{p,i}}
\mathrm{, \quad with \quad} C_d = \left( 0.63+0.48\sqrt{V_r/\mathrm{Re}} \right),
  \label{eqn:beta-i}
\end{equation}
where the particle Reynolds number Re is defined as:
\begin{equation}
  \mathrm{Re} = \rho_i d_{p,i} |\xmbs{u_{p,i}} - \xmbs{U_{f,i}}|,
  \label{eqn:p-re}
\end{equation}
the $V_r$ is the correlation term:
\begin{equation}
  V_r = 0.5\left( A_1 - 0.06\mathrm{Re}+\sqrt{(0.06\mathrm{Re})^2 +
  0.12\mathrm{Re}(2A_2 - A_1)+A_1^2} \right),
  \label{eqn:drag-vr}
\end{equation}
with
\begin{equation}
  A_1 = \varepsilon_f^{4.14}, \quad
  A_2 =
  \begin{cases}
    0.8\varepsilon_f^{1.28} & \quad \text{if } \varepsilon_f \le 0.85, \\
    \varepsilon_f^{2.65}    & \quad \text{if } \varepsilon_f > 0.85.\\
  \end{cases}
  \label{eqn:drag-A}
\end{equation}
The lift force on a spherical particle is modeled in \textit{SediFoam}
as~\citep{saffman65th,rijn84se1}:
\begin{equation}
  \mathbf{f}_{i}^{lift} = C_{l} \rho_f \nu^{0.5} D^{2} \left( \mathbf{u}_{p,i} - \mathbf{U}_{f,i}
  \right) \boldsymbol{\times} \nabla \mathbf{U}_{f,i},
  \label{eqn:particleLift}
  \end{equation}
where $\boldsymbol{\times}$ indicates the cross product of two vectors, and $C_{l} = 1.6$ is the
lift coefficient.
In the present study, the Basset history force and added mass force are not considered. The
calculation of Basset history force requires the integral of the effect of fluid flow on the
particle.  Therefore, this would increase the computational resources significantly. For the added
mass force, we observed that the simulation would be numerically unstable when the numbers of the
layers of sediment particles are large. Hence, the added mass force is not considered in the present
simulations.

\subsection{Diffusion-Based Averaging}

The Eulerian fields $\varepsilon_s$, $\xmb{U}_s$, and $\xmb{F}^{fp}$ in Eq.~(\ref{eq:NS}) are
obtained by averaging the information of Lagrangian particles. The diffusion equations are used to
obtain the continuum Eulerian fields of \( \varepsilon_s \), \( \xmb{U}_s \), and \( \xmb{F}^{fp} \)
by averaging the discrete particle data. The merits of the diffusion-based averaging approach
include (1) sound theoretical foundation, (2) unified treatment of interior and near-boundary
particles, (3) guaranteed conservation of relevant physical quantities, (4) easy implementation in
CFD solvers with arbitrary meshes, and (5) easy parallelization. Here, the averaging process of
$\varepsilon_s$ is discussed as an example. In the first step, the particle volumes at each CFD cell
are obtained. Then, the solid volume fraction for cell \(k\) is calculated by dividing the total
particle volume by the volume of this cell \(V_{c, k}\):
\begin{equation}
  \varepsilon_{s,k}(\mathbf{x},\tau = 0)
 = \frac{\sum_{i=1}^{n_{p, k}} V_{p, i}}{V_{c, k}} , \label{eq:pcm-k} 
\end{equation}
where \(n_{p, k}\) is the number of particles in cell \(k\). With the initial condition in
Eq.~(\ref{eq:pcm-k}), a diffusion equation for $\varepsilon_s(\mathbf{x}, \tau)$ is solved
to obtain the continuum Eulerian field of $\varepsilon_s$: \begin{equation} \frac{\partial
  \varepsilon_s}{\partial \tau} = \nabla^2 \varepsilon_s
\label{eq:diffusion-c}
\end{equation}
where \(\nabla^2 \) is the Laplacian operator and \(\tau\) is pseudo-time. It has been established
in our previous work~\citep{sun14db2,sun14db1} that the results obtained by
Eq.~(\ref{eq:diffusion-c}) are equivalent with Gaussian kernel based averaging with bandwidth $b =
\sqrt{4\tau}$. Similarly, the smoothed \(\xmb{U}_s\) and \(\xmb{F}^{fp}\) fields can be obtained by
using this approach.

\section{Implementations and Numerical Methods}
\label{sec:num-method}
The hybrid CFD--DEM solver \textit{SediFoam} is developed based on two state-of-the-art open-source
codes in their respective fields, i.e., a CFD platform OpenFOAM (Open Field Operation and
Manipulation) developed by \citet{openfoam} and a molecular dynamics simulator LAMMPS (Large-scale
Atomic/Molecular Massively Parallel Simulator) developed at the Sandia National
Laboratories~\citep{lammps}.  At each time step, the fluid and particle equations are solved
individually by the CFD and DEM module. Before solving the fluid equations, the information of the
sediment particles from the DEM module is transferred to the CFD module. Then, an averaging
procedure is performed to ensure the smoothness of the Eulerian fields reconstructed based on the
information of Lagrangian particles.  After solving the fluid equations, the fluid-particle
interaction force of each particle is updated.  Then the DEM module evolves the motion of Lagrangian
particles. An interface is implemented for OpenFOAM and LAMMPS to transfer the particle information
based on OpenMPI. The code is open-source and is available at https://github.com/xiaoh/sediFoam.
Detailed information of the implementations is discussed in our previous paper~\citep{sun2016sedi}.

  It should be noted that the open-source solver \textit{CFDEM} is also a coupled solver of OpenFOAM
  and LAMMPS~\citep{goniva09tf,schmeeckle14ns}. However, \textit{SediFoam} focuses on sediment
  transport problems with several features:
  \begin{enumerate}
    \item The volume fraction terms are not considered in sediment transport applications in
      \textit{CFDEM}~\citep{furbish13ap,schmeeckle14ns}; whereas \textit{SediFoam} first considered
      the volume fraction terms. Since theses terms are considered in \textit{SediFoam}, the mass
      conservation is guaranteed. This is because \textit{SediFoam} uses the diffusion-based
      averaging algorithm and can obtain smooth Eulerian fields from Lagrangian particles on
      arbitrary meshes.
    \item Pressure gradient force, fluid viscous force, lift force, lubrication force and added mass
      force, which are not negligible in sediment transport applications, are implemented in
      \textit{SediFoam}. However, in \textit{CFDEM}, only pressure gradient force and fluid
      viscous forces are considered.
    \item The parallel efficiency of \textit{SediFoam} is very good when using up to 512 processors
      on simulations of 40 million sediment particles. This is because the parallel efficiency of
      the interface between OpenFOAM and LAMMPS is satisfactory.
  \end{enumerate}

The fluid equations in~(\ref{eq:NS}) are solved in OpenFOAM with the finite volume method
\citep{jasak96ea}. PISO (Pressure Implicit Splitting Operation) algorithm is used to prevent
velocity--pressure decoupling~\citep{issa86so}. A second-order central scheme is used for the
spatial discretization of convection terms and diffusion terms. Time integrations are performed with
a second-order implicit scheme. In the averaging procedure, the diffusion equations are solved on
the same mesh as the CFD mesh~\citep{capecelatro13ae,sun14db1}. A second-order central scheme is
used for the spatial discretization of the diffusion equation; a second-order implicit scheme is
used for the temporal integration.  To model the collision between the sediment particles, the
contact force between sediment particles is computed with a linear spring-dashpot model. In this
model, the normal push-back force for two overlapping particles is linearly proportional to the
overlap distance; the damping force is proportional to the relative velocity between two overlapping
particles~\citep{cundall79}. This linear spring-dashpot model is proven to be adequate in the
simulation of particle-laden flows~\citep{kidanemariam14dn,gupta15vv}.

\section{Results}
\label{sec:simulations}

Setups of the numerical simulations performed here follow the experimental study of sediment
transport in an asymmetric oscillatory flow~\citep{od04co,od04ft}. The purposes of the numerical
simulations are (1) to validate the capability of the CFD--DEM model under oscillatory flow actions,
and (2) to reveal the physical insights of the variation of micromechanic variables at different
phases in the oscillatory flow cycle. In Section~\ref{sec:run-valid}, the results obtained in the
numerical simulations are compared to the experimental results, which aims to demonstrate that
CFD--DEM is capable of predicting the sediment transport in oscillatory sheet-flows. In
Section~\ref{sec:run-micro}, the study of the micromechanics of sediment bed is detailed. The
Voronoi volume fraction, the coordination number, and the inter-particle contact force are
presented, which aims to compare the micromechanics of sediment bed at different phases. In
addition, the effect of the particle size on these micromechanic variables is demonstrated.

\subsection{Numerical validation}
\label{sec:run-valid}
\begin{figure}[htbp]
  \centering
  \includegraphics[natheight = 500, natwidth = 700,width=0.8\textwidth]{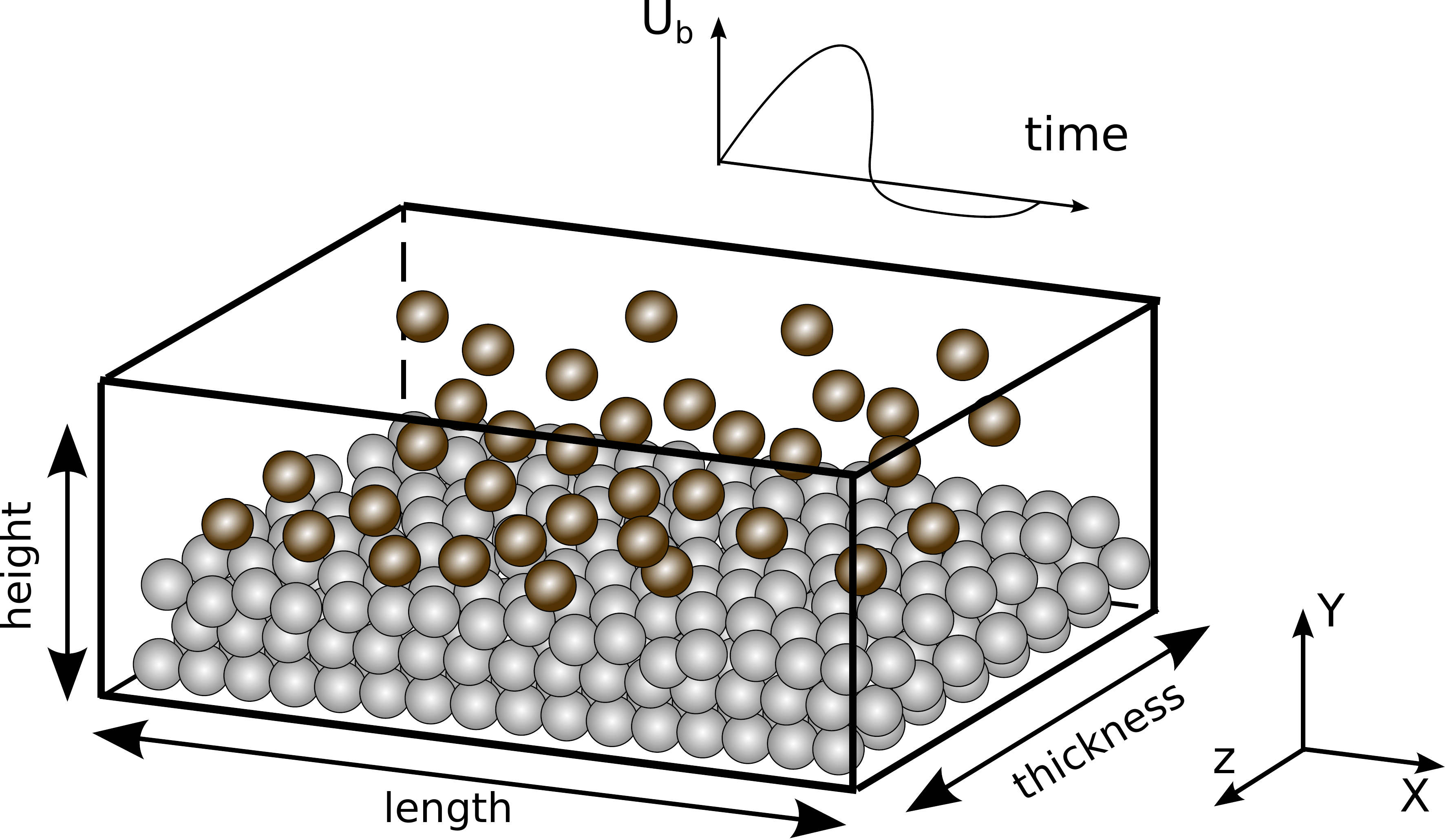}
  \caption{The geometry of the computational domain in the numerical simulations. The grey
  ``immobile'' particles are fixed at the bottom; the brown particles are movable. For clarity, only
  a few mobile particles are shown here for illustration purpose. The simulations used 250,000 and
  108,000 particles. The particle sizes are not to scale.}
  \label{fig:layout}
\end{figure}

The numerical tests are performed using a periodic channel according to the
literature~\citep{od04co,od04ft,malarkey09mo}. The geometry of the computational domain and the
coordinates system are shown in Figure~\ref{fig:layout}. The Cartesian coordinates $x$, $y$, and $z$
are aligned with the streamwise, vertical, and lateral directions, respectively. The boundary
conditions in both $x$- and $z$-directions are periodic. The no-slip wall boundary condition is
applied at the bottom while free-slip condition is applied on the top.  The three fixed layers are
arranged in hexagonal packing according to~\citep{kempe14ot}. Since the velocity of the particles at
the bottom is very small, the pattern of the fixed particle does not significantly influence the
results.  The CFD mesh is refined in the vertical ($y$-) direction at the particle-fluid interface
to resolve the flow structure in the boundary layer. The top of the initial bed corresponds to $y =
0$. The physical parameters used are detailed in Table~\ref{tab:parameters}.  Based on previous
studies on sediment transport in unidirectional flow using
CFD--DEM~\citep{furbish13ap,schmeeckle14ns}, the size of the computational domain is chosen to be
100--240$d_p$ in streamwise direction and 60--120$d_p$ in lateral direction. Since the size of the
computational domain in the present study is within the range of previous studies (see
Table~\ref{tab:parameters}), the computational domain used is adequate. In addition, the numbers of
sediment particles range from 24,000 to 330,000 in previous tests in unidirectional
flow~\citep{furbish13ap,schmeeckle14ns}. Since the numbers of sediment particles in the present
study are 256,000 and 108,000, the numbers of sediment particle are also adequate.  It should be
noted that the mesh resolutions are different in the cases because the diameters of the particles
used are different. In the present simulations, the sediment particles are arranged in hexagonal
lattice, and the size of the CFD mesh is $2d_p/\sqrt{3}$ in streamwise direction and $2d_p$ in
lateral direction.  This is to avoid the variation of solid volume fraction $\varepsilon_s$ on the
bottom CFD cells for the uniform distributed bottom particles. Since the computational domains in
the cases are similar (although not consistent), the mesh resolutions are different. From
mesh-independent study by~\cite{schmeeckle14ns}, the mesh resolution in the present study is
adequate to resolve the turbulent eddies at the boundary layer.  The experimental measurements are
performed using coarse, medium and fine sands. In this work, we focus on the simulations of coarse
and medium sands.  The simulation of fine sand case is not performed because the computational costs
are unaffordable when the erosion depth is approximately $100d_p$. The height of the domain in the
numerical simulation is 75~mm, which is smaller than the channel height 750~mm in the experiment.
This reduction of the computational domain is to limit the computational costs. It is noted that the
sediment particle used in the experiment is well-sorted, but uniform-sized particles are used in the
present simulations. This is because the detailed particle size distribution is not given. Moreover,
the results obtained in previous studies using uniform-sized particle are
satisfactory~\citep{malarkey09mo}.

The stiffness, the restitution coefficient, and the friction coefficient are also detailed in
Table~\ref{tab:parameters}.  In the present simulations, the restitution coefficient $e$ is 0.01
according to~\cite{schmeeckle14ns}, which is smaller than that used by~\cite{drake01dp}. This is due
to the fact that the effective restitution coefficient of coarse and medium particles (Stokes number
$ < 10$) in water is very small when the lubrication layer is formed between the contact
particles~\citep{Kempe_12_CM}. The selection of the friction coefficient in the present study
according to~\cite{kidanemariam14dn}. However, the results are not sensitive to the restitution
coefficient and friction coefficient in the present simulations.  The typical contact time is
$4.6\times10^{-4}$~s for coarse sand and $2.0\times10^{-4}$~s for medium sand. The DEM step in the
present study is $2.0\times10^{-6}$~s, which is smaller than 1/50 of the contact time to avoid
particle inter-penetration~\citep{sun07ht}.

\begin{figure}[htbp]
  \centering
  \includegraphics[natheight = 500, natwidth = 700,width=0.8\textwidth]{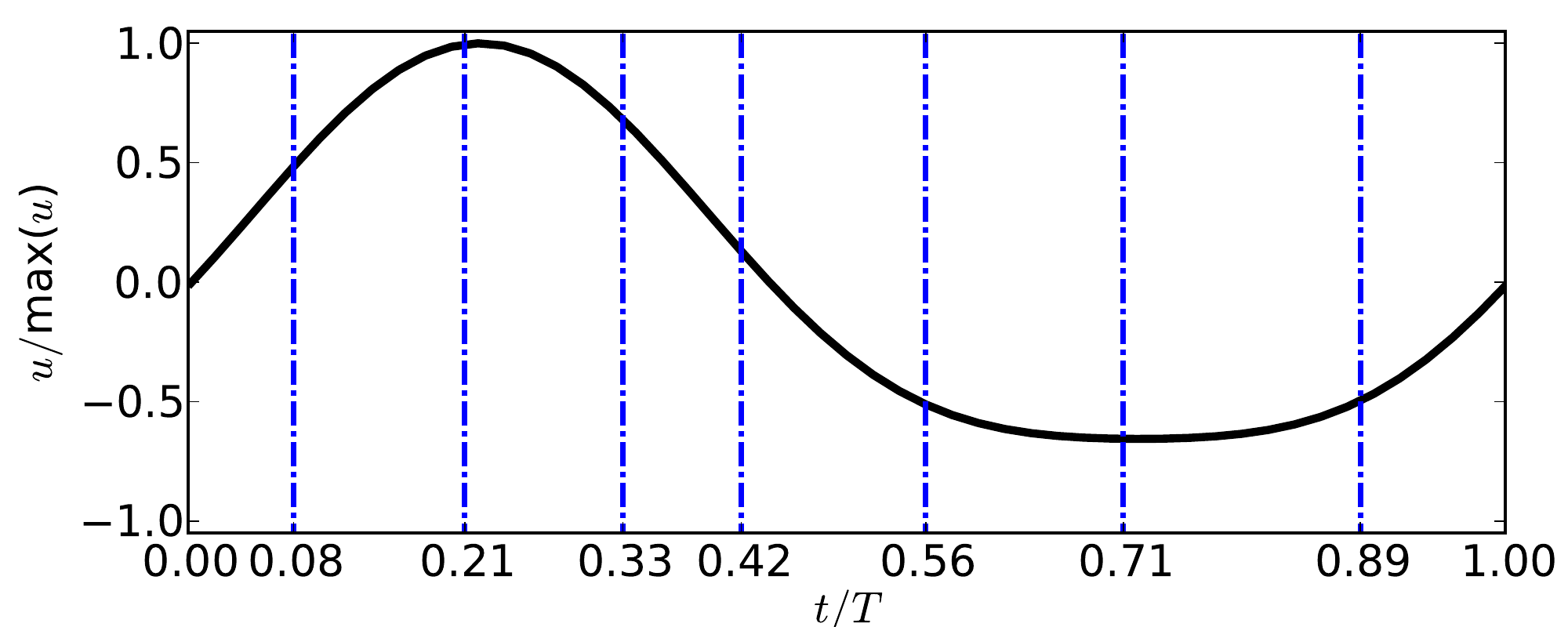}
  \caption{Time history of the mean flow velocity. Vertical lines indicate eight phases at which the
  results obtained from the numerical simulations are compared to the experiments.}
  \label{fig:wave-his}
\end{figure}

The asymmetric sheet flow velocity is shown in Figure~\ref{fig:wave-his}. The time history of this
oscillatory flow is:
\begin{equation}
  u(t) = u_1\sin(\omega t) - u_2 \cos(2\omega t),
  \label{eq:wave-velocity}
\end{equation}
where $\omega = 2\pi/T$; $u_1$ and $u_2$ determine the magnitude and asymmetry of the wave,
respectively.  According to~\cite{od04ft}, when the flow velocity is positive, it is
onshore-directed; whereas the flow is offshore-directed when the velocity is negative. This second
order Stokes wave is typical of natural surf zone in the production of sediment
transport~\citep{drake01dp}.  The fluid flow is driven by an oscillatory pressure gradient $p_d(t)$,
which is the time derivative of flow velocity:
\begin{equation}
  p_d(t) = u_1\omega\cos(\omega t) + 2u_2\omega\sin(2\omega t).
  \label{eq:wave-pressure}
\end{equation}
The flow velocity and sediment flux obtained in the present simulation are validated by using the
experimental data at different phases of the oscillatory flow cycle, which are indicated as the
vertical line in Figure~\ref{fig:wave-his}. 

\begin{figure}[htbp]
  \centering
  \subfloat[]{
  \includegraphics[natheight = 500, natwidth = 700,width=0.45\textwidth]{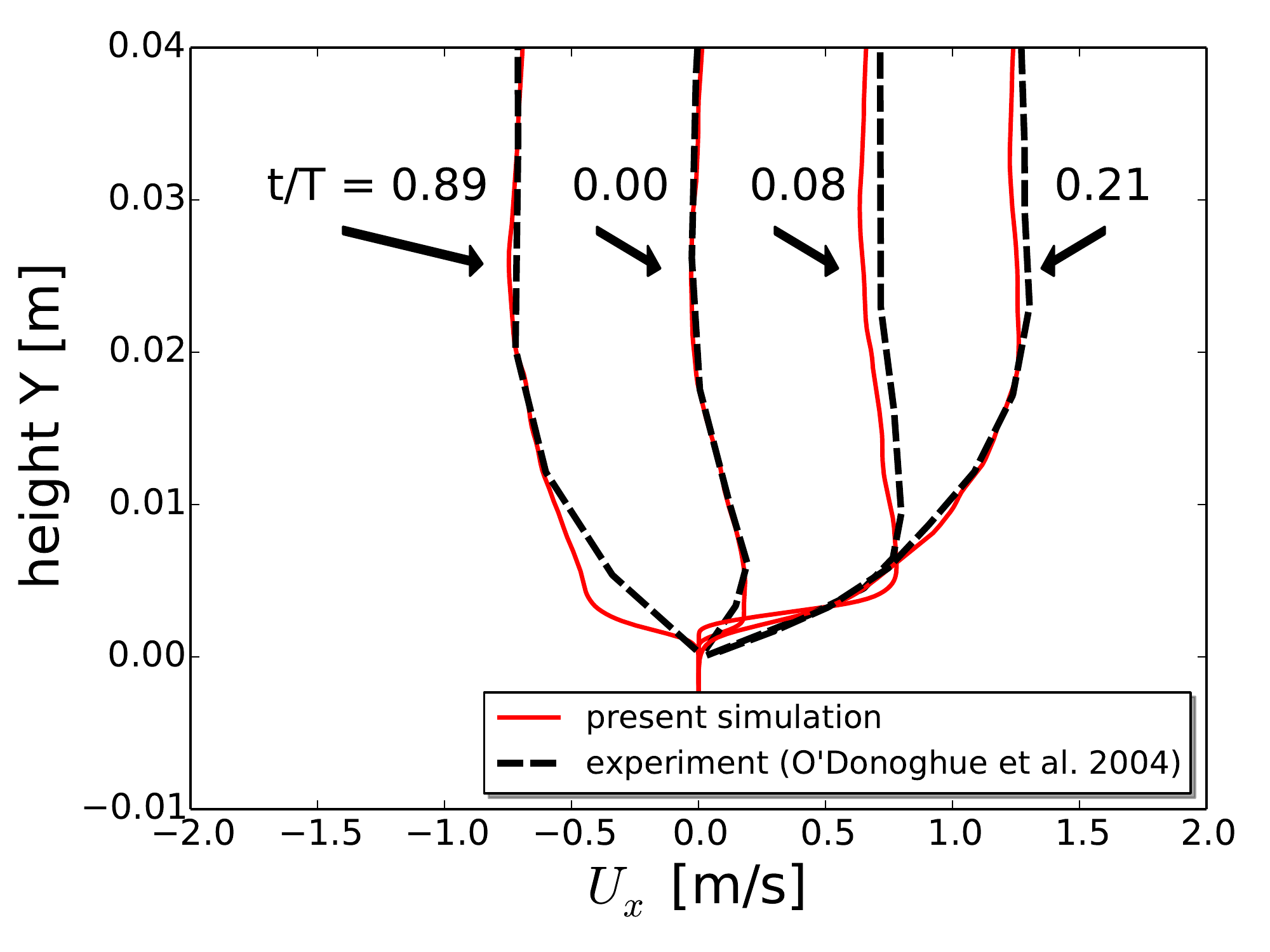}
  }
  \subfloat[]{
  \includegraphics[natheight = 500, natwidth = 700,width=0.45\textwidth]{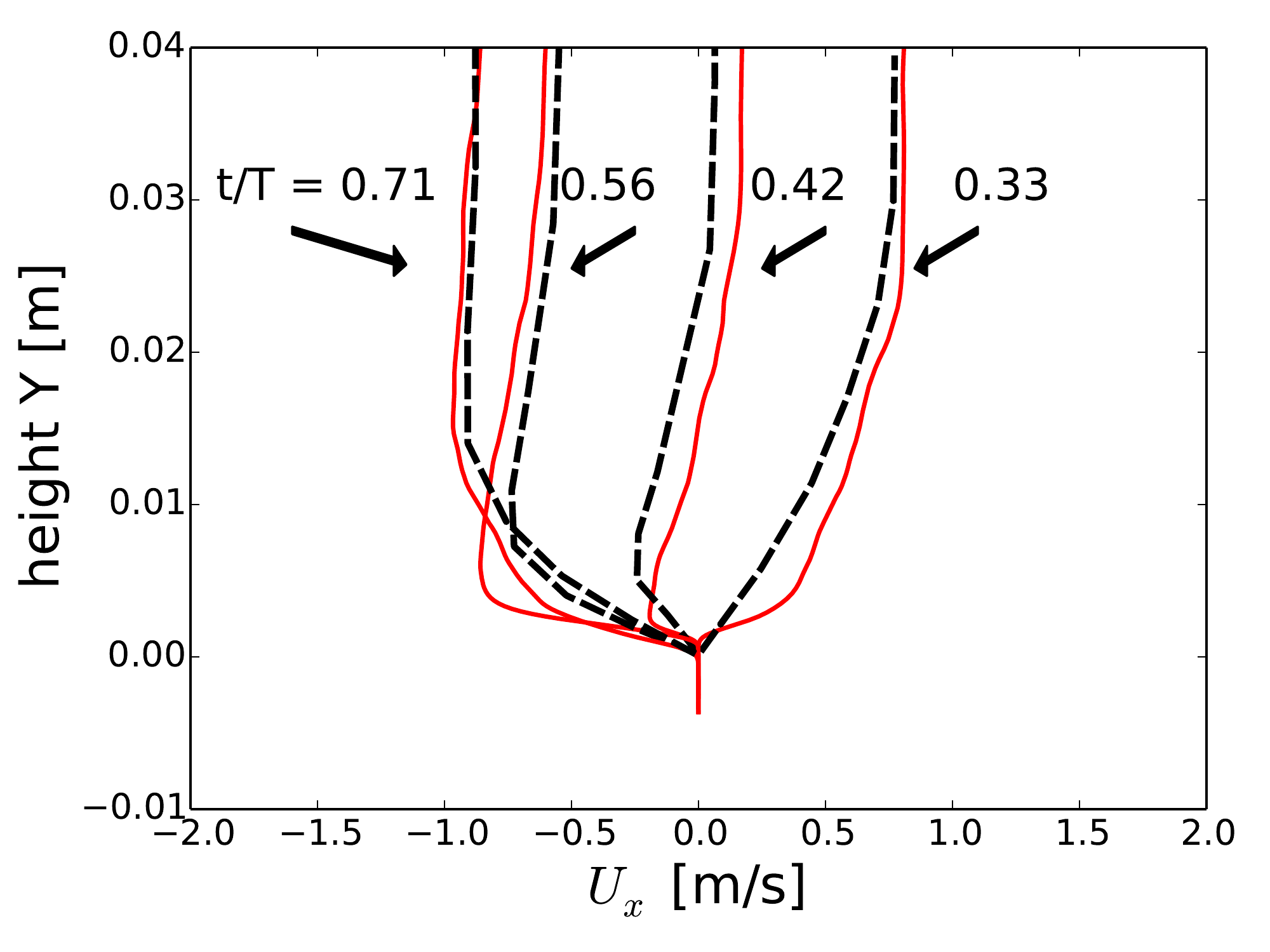}
  }
  \caption{Horizontally-averaged streamwise flow velocity for Case 1 at different phases in the
    oscillatory flow cycle. To show each velocity profile clearly, the profiles are splited
    according to Figure~6 in~\cite{od04ft}.}
  \label{fig:Ub-inst}
\end{figure}

The profiles of horizontally averaged flow velocity in the streamwise direction obtained in the
present simulations are shown in Figure~\ref{fig:Ub-inst}. It can be seen in the figure that the
flow velocity profiles obtained by using CFD--DEM agree with the experimental data at all the phases
examined in the oscillatory cycle.  The height of the computational domain is smaller than that in
the experiment, but it is sufficient to capture the bulk physics of the fluid flow. In addition to
the flow velocity, the sediment flux is also examined to validate the numerical simulation. The
sediment flux $\varphi$ is defined as $\varphi = \varepsilon_s U_{f,x}$, where $U_{f,x}$ is the
streamwise component of the fluid velocity.  Although the particle velocity $U_{s,x}$ should be used
in the definition of sediment flux~\citep{schmeeckle14ns}, the present study uses the fluid velocity
$U_{f,x}$ to be consistent with the definition in the experimental studies~\citep{od04co,od04ft}.
This is attributed to the fact that the particle velocity is difficult to measure in the oscillatory
flow experiment, and thus the fluid velocity is used instead.  Considering the terminal velocity of
the sediment particle is very small compared with the flow velocity, the differences between
horizontally averaged fluid velocity and particle velocity are not significant. 

\begin{figure}[htbp]
  \centering
  \subfloat[$t/T$ = 0]{
  \includegraphics[width=0.24\textwidth]{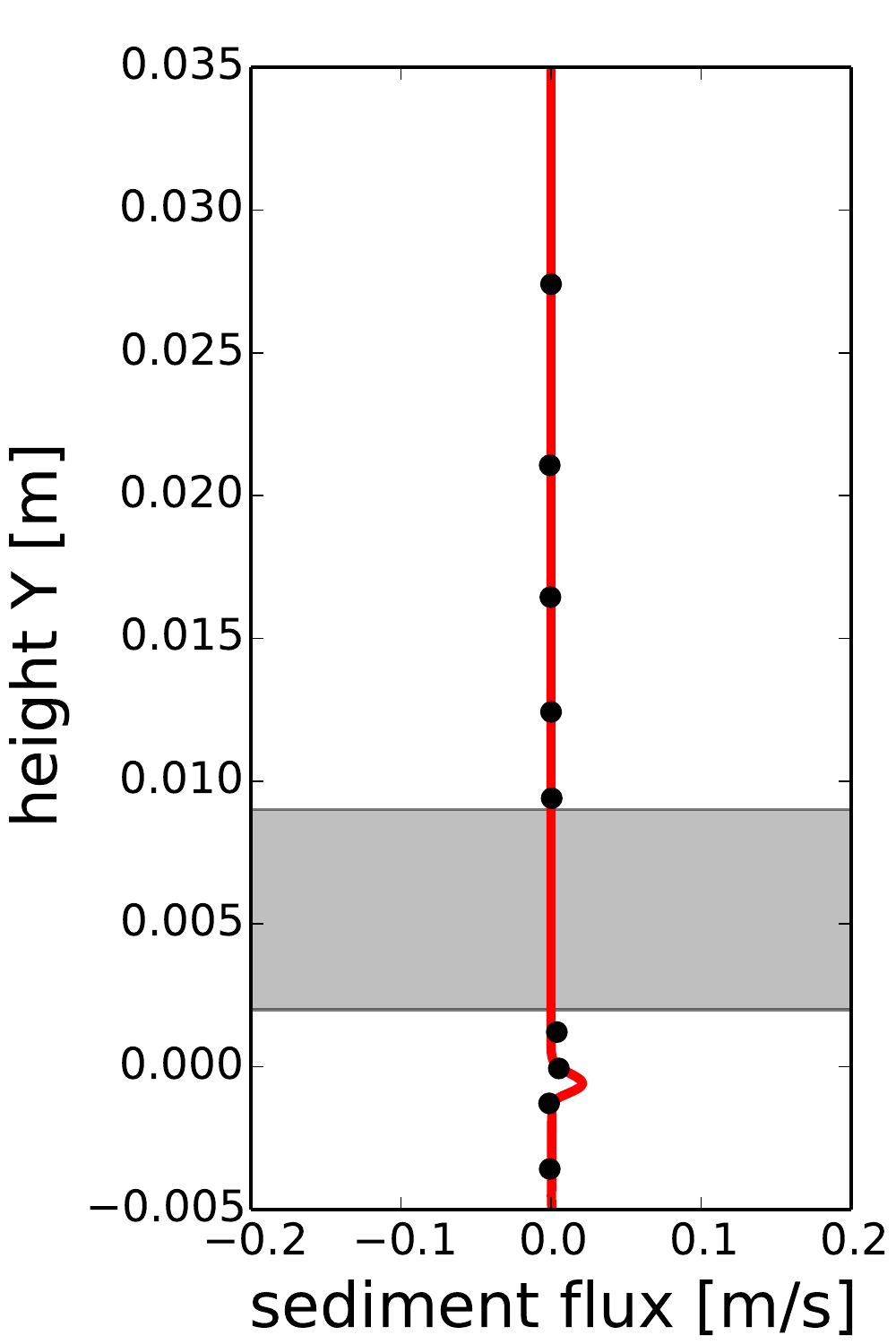}
  }
  \hspace{-0.19in}
  \subfloat[$t/T$ = 0.08]{
  \includegraphics[width=0.24\textwidth]{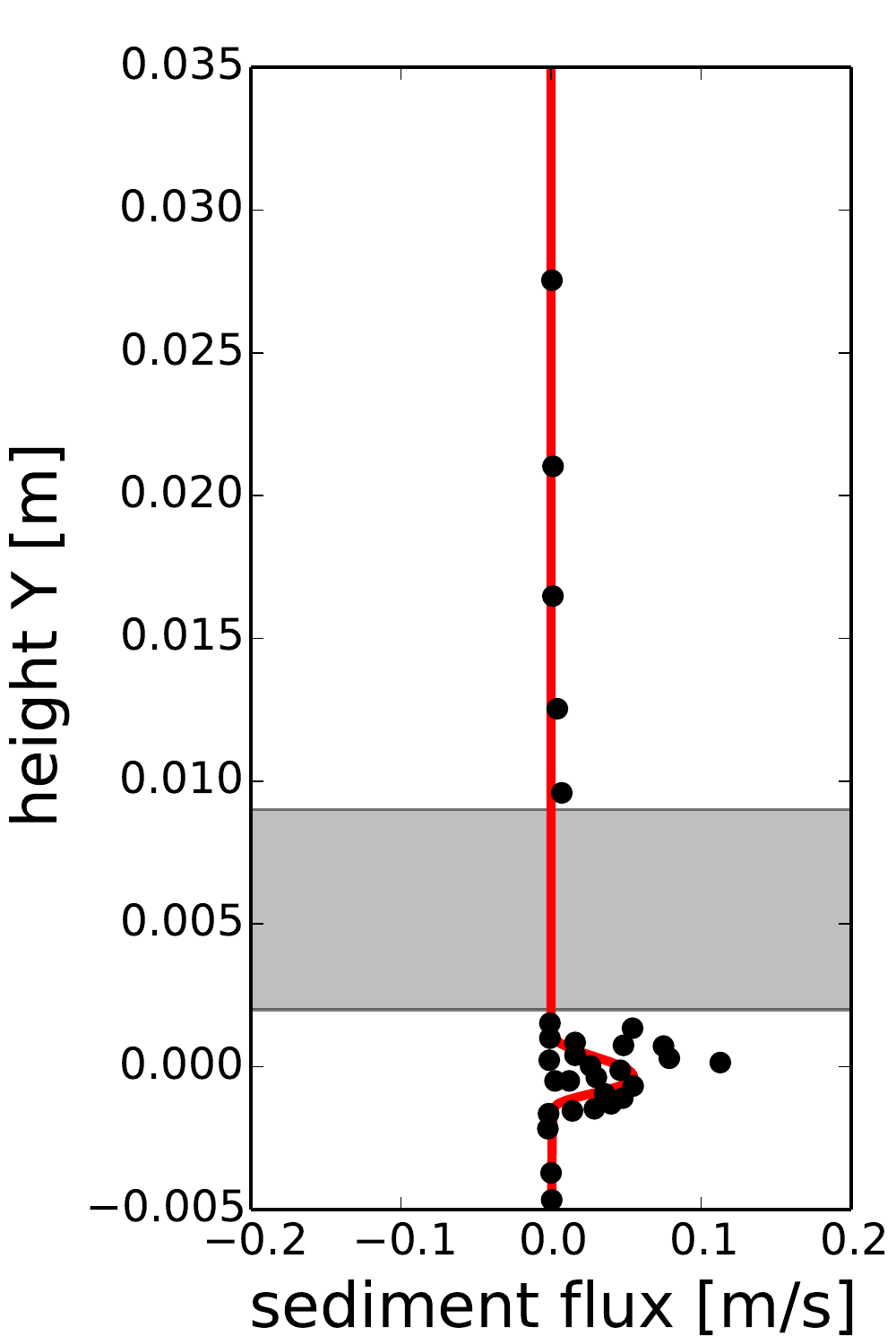}
  }
  \hspace{-0.19in}
  \subfloat[$t/T$ = 0.21]{
  \includegraphics[width=0.24\textwidth]{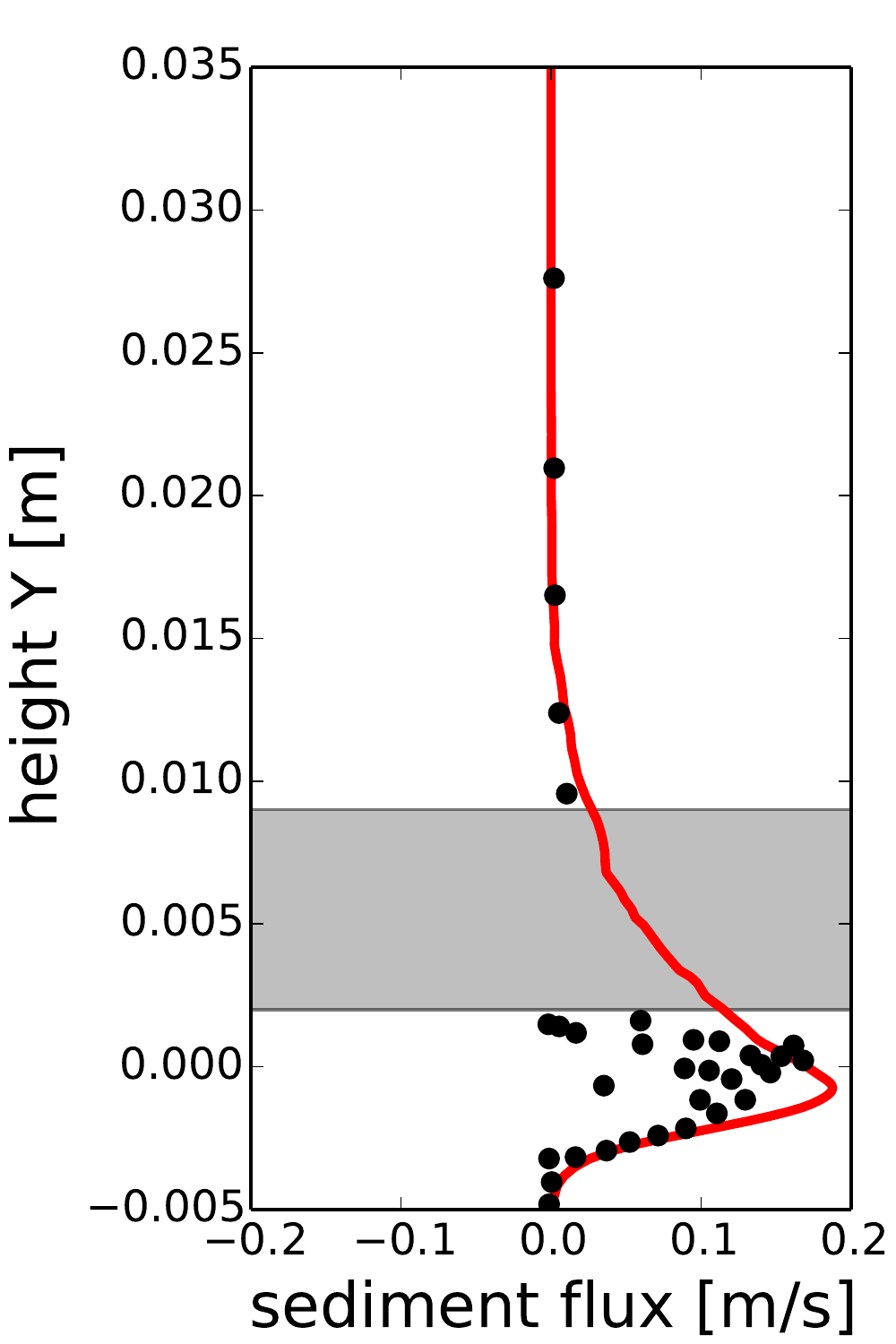}
  }
  \hspace{-0.19in}
  \subfloat[$t/T$ = 0.33]{
  \includegraphics[width=0.24\textwidth]{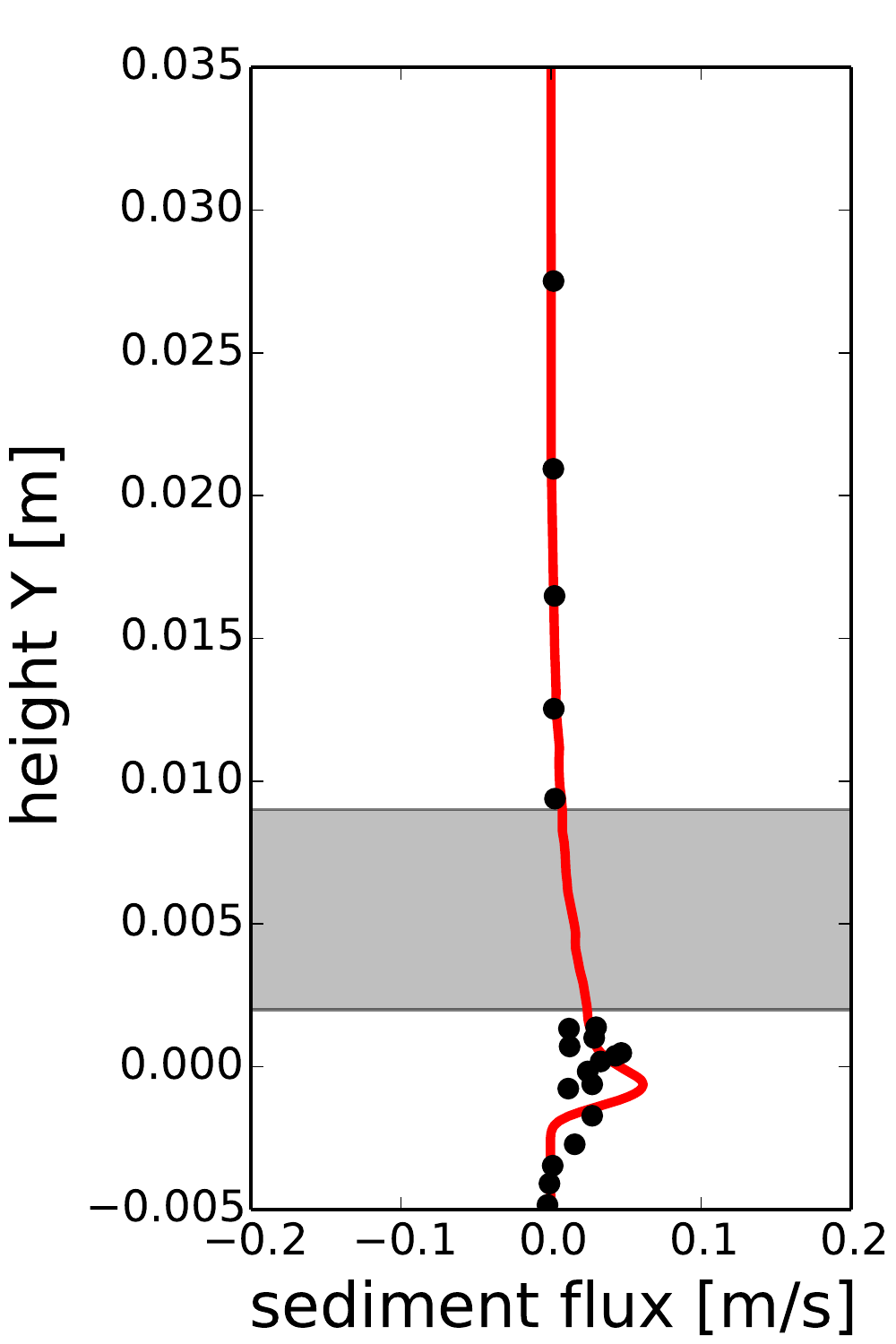}
  }
  \vspace{0.1in}
  \subfloat[$t/T$ = 0.42]{
  \includegraphics[width=0.24\textwidth]{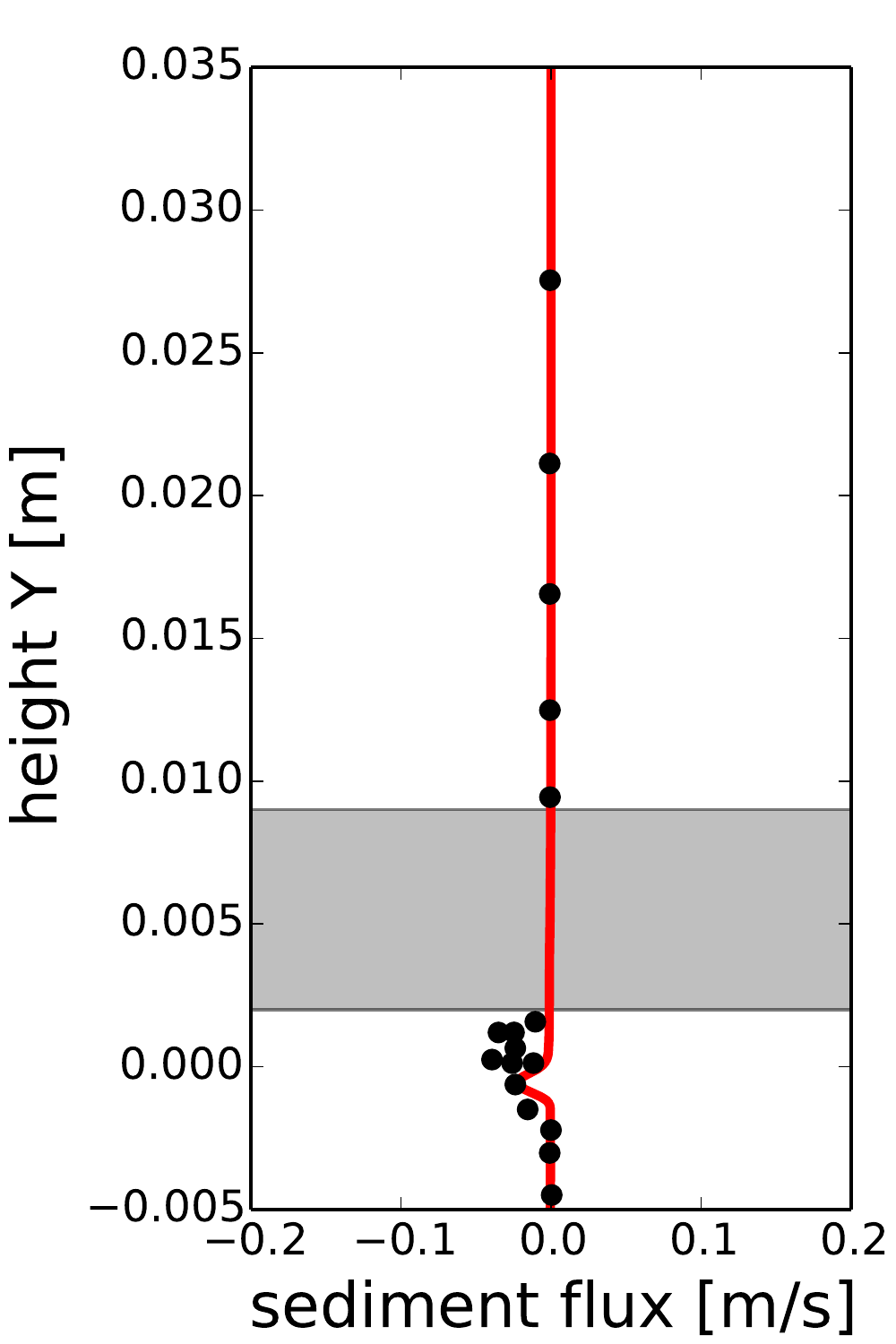}
  }
  \hspace{-0.19in}
  \subfloat[$t/T$ = 0.56]{
  \includegraphics[width=0.24\textwidth]{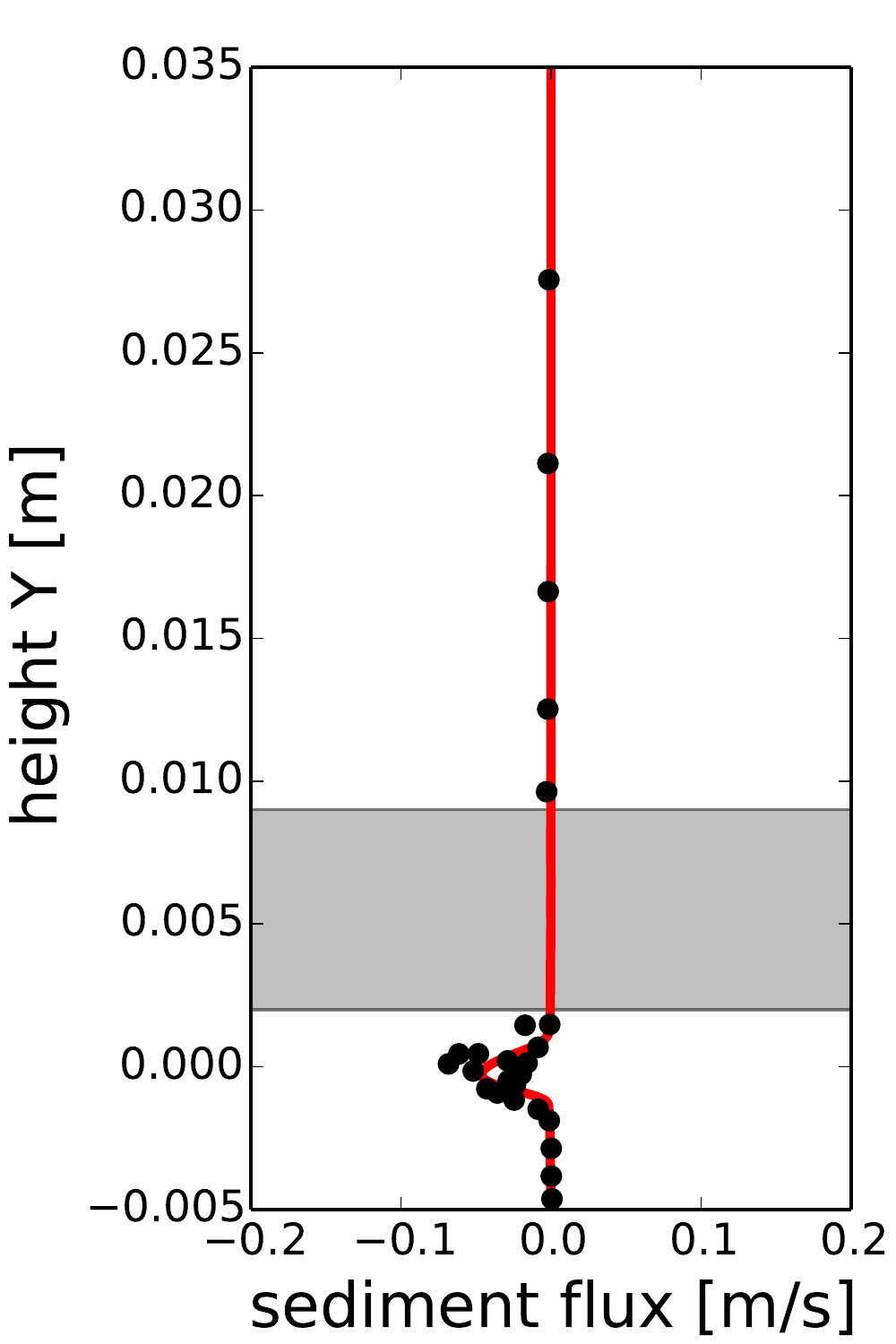}
  }
  \hspace{-0.19in}
  \subfloat[$t/T$ = 0.71]{
  \includegraphics[width=0.24\textwidth]{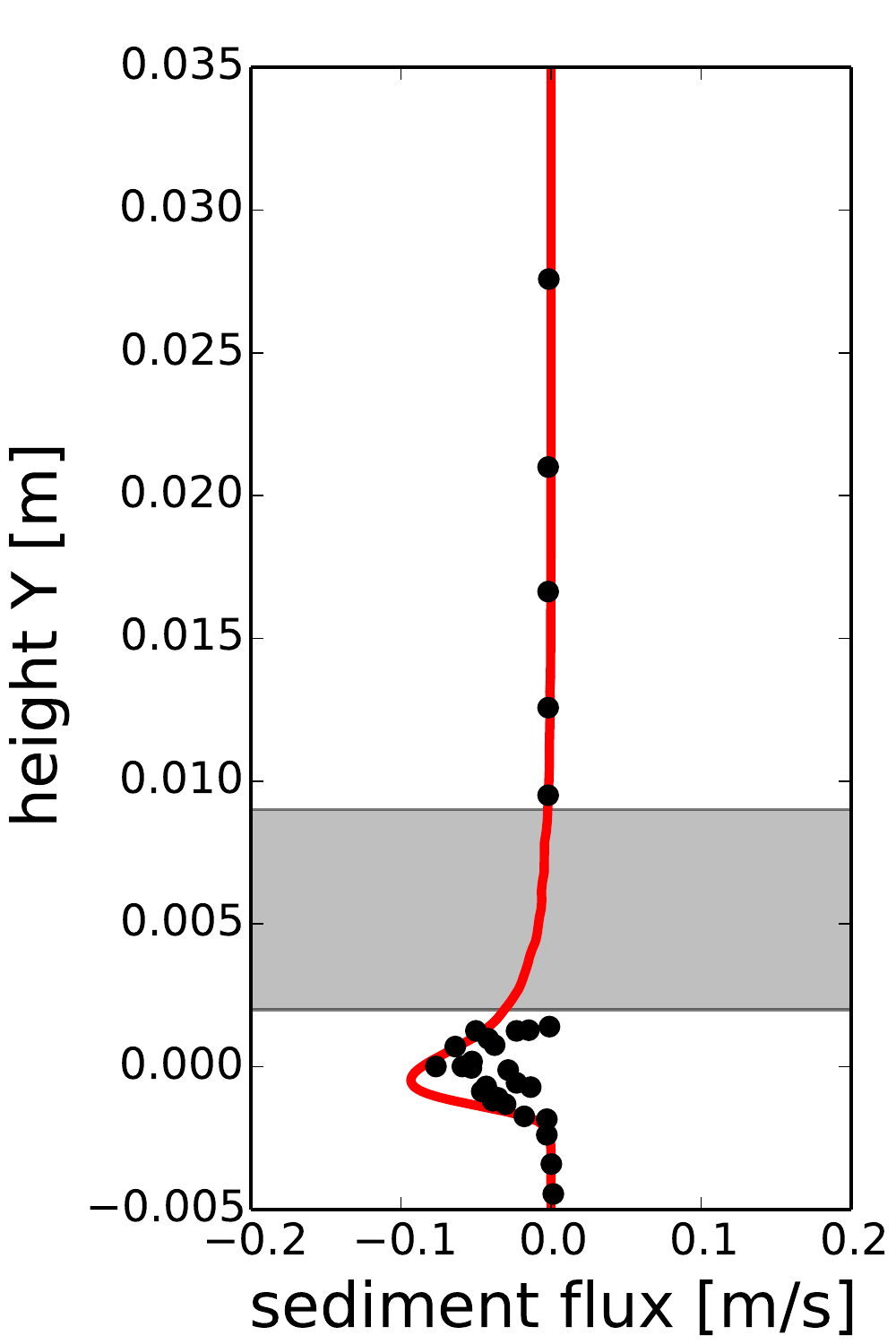}
  }
  \hspace{-0.19in}
  \subfloat[$t/T$ = 0.89]{
  \includegraphics[width=0.24\textwidth]{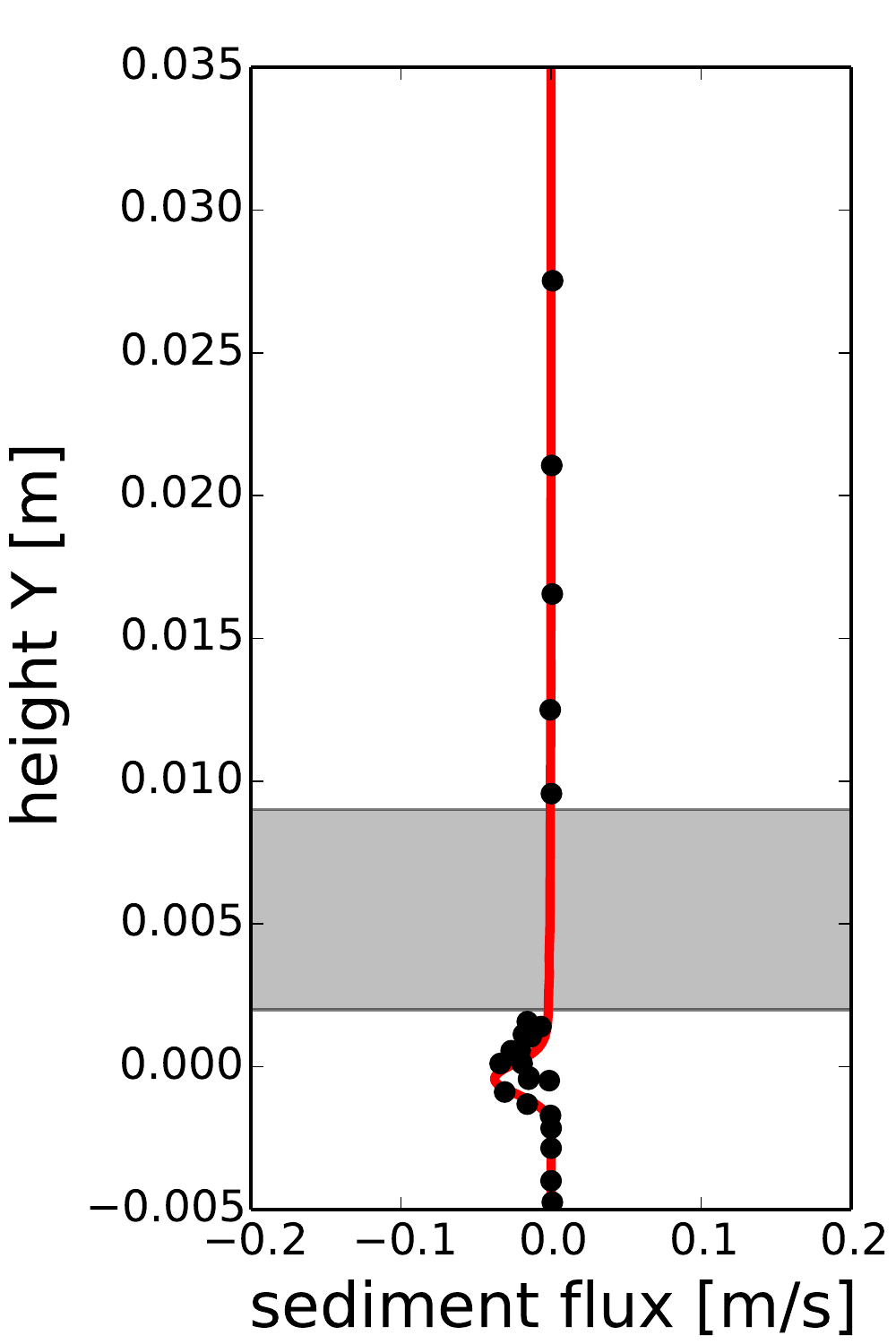}
  }

  \caption{Horizontally-averaged streamwise sediment flux for Case 1 at different phases. The red
    solid lines are the results obtained by using CFD--DEM; the black dots are the experimental
  measurements. It is noted that there is a gap (highlighted in shade) in the experimental data
  between conductivity concentration probes ($h < 2$~mm) and suction samplers ($h > 10$~mm). The
  initial surface of the sediment bed is $y = 0$.}
  \label{fig:Q-inst}
\end{figure}

Figure~\ref{fig:Q-inst} shows that the vertical profiles of sediment flux obtained from the
numerical simulations are consistent with the experiment results.  When measuring high concentration
regions in the experiments, CCM probes can only be applied at the bottom (concentration higher than
6\%, which is located within 2~mm above the initial sediment bed) to provide good concentration
measurements. To measure low concentration regions, the suction samplers can produce a significant
lowering of the mean bed level (from 10 to 30~mm) in their vicinity, and thus are located
approximately 10~mm above the bed to reduce its influence on the seabed ($h > 10$~mm).  Because of
the limitations of the experimental measurements, there is a gap of the order of 10~mm profiles of
sediment flux. It can be seen in Figure~\ref{fig:Q-inst} that the agreement between the experimental
results and numerical simulation is satisfactory for both regions measured by CCM probes and suction
samplers. The discrepancy between the results obtained in the experiments and numerical simulations
is attributed to the uncertainty of the measurement and numerical model.  The gap in the
experimental measurement is bridged since the concentration and velocity of the moving particles are
easier to obtain in the simulations. This is an advantage of the CFD--DEM approach over the
experimental measurement.  Based on the sediment flux obtained in the simulations, the particles on
the surface of the sediment bed are moving with fluid flow, but those on the bottom are almost
stationary. When the particle number increases in the simulations, only the number of non-moving
particle is increasing and the results are not sensitive to the number of particles. This shows the
number of sediment particles is sufficient in the present simulations.

\begin{table}[!htbp]
  \caption{Parameters used in numerical simulations of sediment transport.}
 \begin{center}
 \begin{tabular}{lcccc}
   \hline
   & \thead{Case 1 \\ (medium sand)} & \thead{Case 2 \\ (coarse sand)} \\
   \hline
   bed dimensions & & \\
   \hline
   \qquad width $(L_x)$                 & 139$d_p$ (37.4~(mm))     & 104$d_p$ (47.8~(mm))  \\
   \qquad height $(L_y)$                & 278$d_p$ (75~(mm))       & 163$d_p$ (75~(mm))    \\
   \qquad transverse thickness  $(L_z)$ &  80$d_p$ (21.6~(mm))     &  60$d_p$ (27.6~(mm))  \\
   \hline
   mesh resolutions                         &       &       \\ 
   \hline
   \qquad width $(N_x)$                     & 80    & 60    \\
   \qquad height $(N_y)$                    & 180   & 180   \\
   \qquad transverse thickness $(N_z)$      & 40    & 30    \\
   \hline
   particle properties  & & \\
   \hline
   \qquad total number                      & 256,000       & 108,000 \\
   \qquad diameter $d_p$                    & 0.27~(mm)     & 0.46~(mm)    \\
   \qquad density $\rho_s$  & \multicolumn{2}{ c }{$2.65\times10^3$~($\mathrm{kg/m^3}$)} \\
   \qquad particle stiffness coefficient    & \multicolumn{2}{ c }{20~(N/m)} \\
   \qquad normal restitution coefficient    & \multicolumn{2}{ c }{0.01} \\
   \qquad coefficient of friction           & \multicolumn{2}{ c }{0.4} \\
   \hline
   fluid properties & & \\
   \hline
   \qquad density $\rho_f$      & \multicolumn{2}{ c }{$1.0\times10^3$~($\mathrm{kg/m^3}$)} \\
   \qquad kinetic viscosity $\nu$   & \multicolumn{2}{ c }{$1.0\times10^{-6}$~($\mathrm{m^2/s}$)} \\
   \hline
   wave properties (see Eq.~(\ref{eq:wave-velocity})) & & \\
   \hline
   \qquad $u_1$                 & \multicolumn{2}{ c }{ 1.06~(m/s) }\\
   \qquad $u_2$                 & \multicolumn{2}{ c }{ 0.22~(m/s) }\\
   \qquad period ($T$)          & \multicolumn{2}{ c }{ 5.0~(s) }\\
   \hline
  \end{tabular}
 \end{center}
 \label{tab:parameters}
\end{table}

\begin{figure}[htbp]
  \centering
  \subfloat[ Case 1 (medium sand) ]{
  \includegraphics[width=0.45\textwidth]{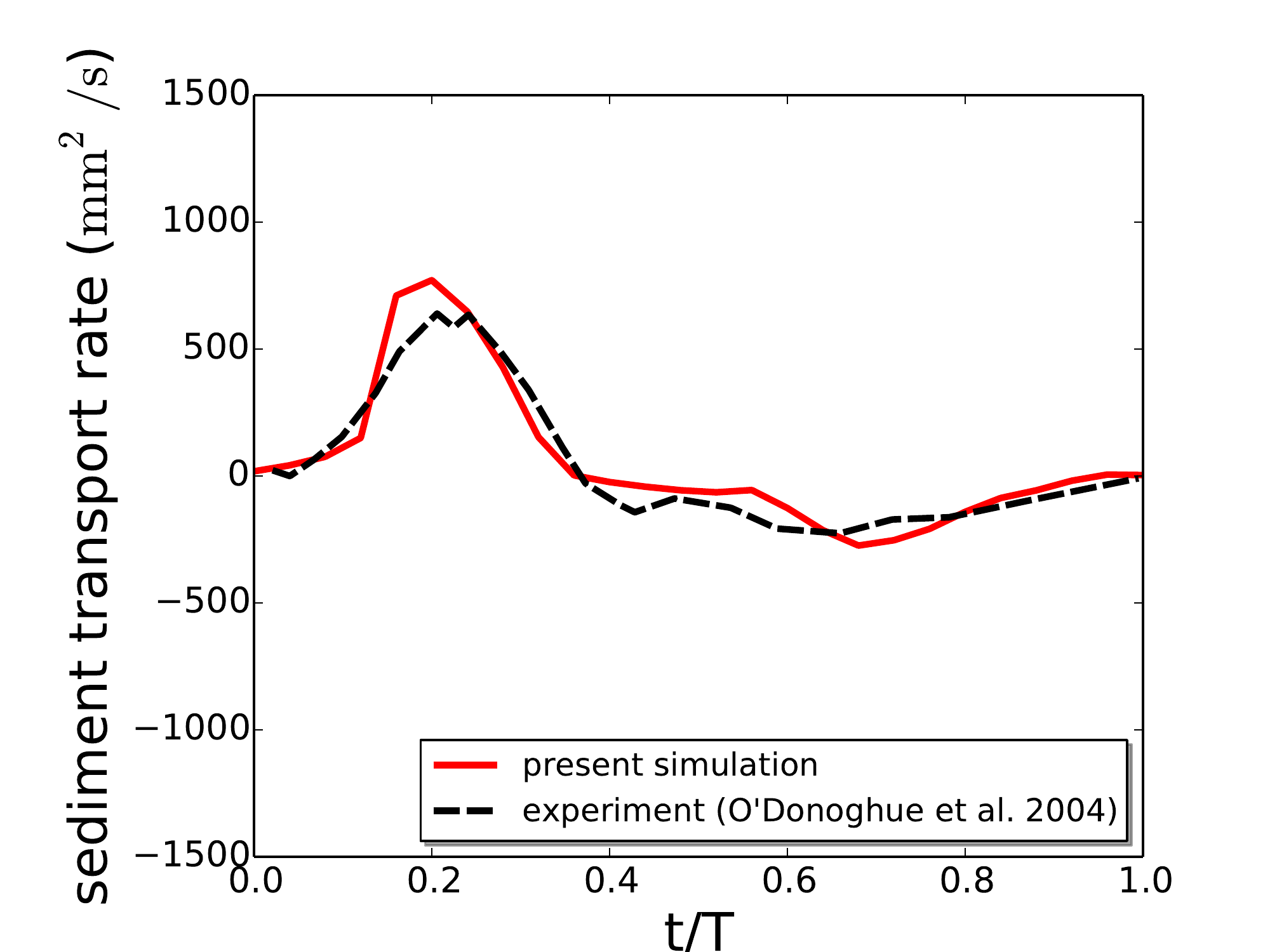}
  }
  \subfloat[ Case 2 (coarse sand) ]{
  \includegraphics[width=0.45\textwidth]{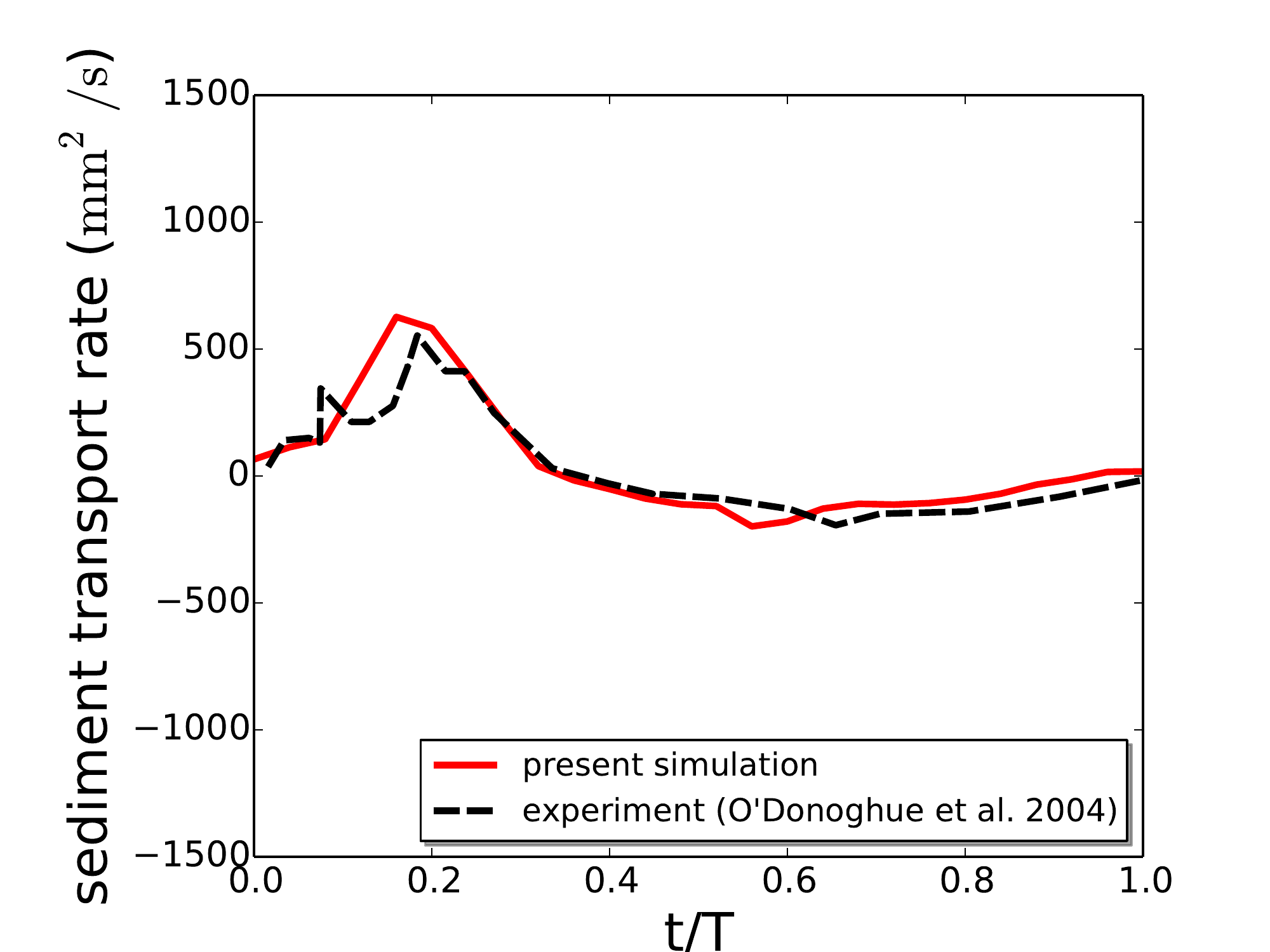}
  }
  \caption{Time-hisotry of the oscillatory sheet-flow sediment transport rate for Cases 1 and 2.}
  \label{fig:net-rate-all}
\end{figure}

Another quantity of interest is the transport rate of the sediment particles, which is defined as
the spacial integral of the sediment flux at the sheet flow layer:
\begin{equation}
  q_{sf} = \int_{-\sigma_e}^{y_s} \varphi \mathrm{d}y,
  \label{eq:q_sf}
\end{equation}
where $\sigma_e$ is the height of the erosion depth; $y_s$ is the height of the sheet flow layer.
According to~\cite{od04co}, the erosion depth is defined according to a 2-step procedure: (1) find
the vertical concentration profile of best fit; (2) draw a tangent line from the inflection point on
the profile, and find the point of intersaction between the concentration profile and $\varepsilon_s
= 0.6$.  However, in the present study, the sediment flux at the bottom is very small and the
integral in Eq.~(\ref{eq:q_sf}) starts from the bottom of the computational domain.  The height of
the sheet flow layer is defined at the location where $\varepsilon_s = 0.08$.  The sediment
transport rates $q_{sf}$ obtained in the CFD--DEM simulations are compared with the experimental
results for Cases 1 and 2. It can be seen in Figure~\ref{fig:net-rate-all} that the sediment
transport rates obtained in the CFD--DEM simulations are consistent with the experimental data in
both cases. CFD--DEM captures the peak at $t/T = 0.21$ and the trough of the sediment transport
rate, which demonstrates the capability of CFD--DEM in the prediction of both onshore and offshore
sediment transport. The time integral of the transport rate in one period is the net sediment
transport rate, which can be used to evaluate the rate of onshore and offshore sediment transport.
The net sediment transport rates of the two cases are 53 and 44 $\mathrm{mm}^2/\mathrm{s}$, which
agree very well with the experimental measurement of 56 and 45
$\mathrm{mm}^2/\mathrm{s}$~\citep{malarkey09mo}.  According to the results, the asymmetric wave
leads to net onshore sediment transport for both cases.

In summary, the overall agreement between the results obtained by the numerical simulation and
experimental measurement is satisfactory in general. This indicates that the CFD--DEM model is
capable of predicting the sediment flux in oscillatory flows.

\subsection{Granular Micromechanics}
\label{sec:run-micro}

The micromechanic variables at different phases in the oscillatory cycle of both cases are compared,
including the Voronoi volume fraction, the coordination number, and the contact force. The volume
fraction and coordination number describe the packing of sediment bed; the total contact force
indicates the particle interaction. For each sediment particle, the Voronoi volume fraction is the
ratio of the volume of this particle and the total volume this particle occupies; the coordination
number is the number of contact from neighboring particles; the total contact force is the resultant
of the contact force from neighboring particles. The purpose of the comparison of micromechanic
variables is to demonstrate the evolution of sediment bed in oscillatory flows. Detailed definitions
of the micromechanic variables and post-processing procedures are discussed in the Appendix.

\begin{figure}[htbp]
  \centering
  \subfloat[ Case 1 (medium sand), vertical profile ]{
  \includegraphics[width=0.45\textwidth]{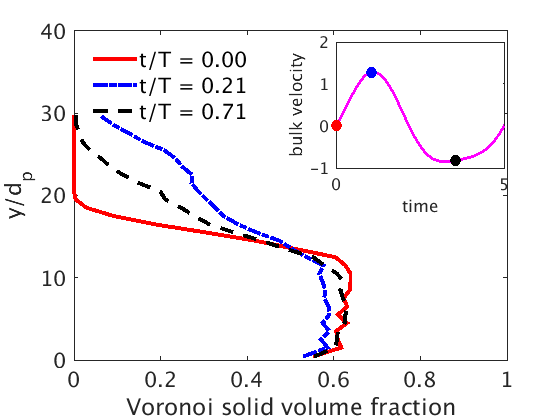}
  }
  \subfloat[ Case 2 (coarse sand), vertical profile ]{
  \includegraphics[width=0.45\textwidth]{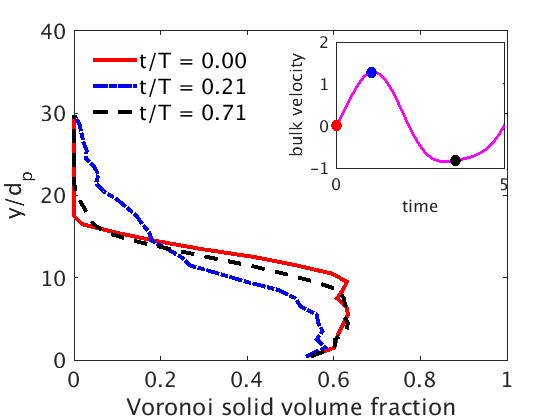}
  }
  \vspace{0.1 in}
  \subfloat[ Case 1 (medium sand), probability density]{
  \includegraphics[width=0.45\textwidth]{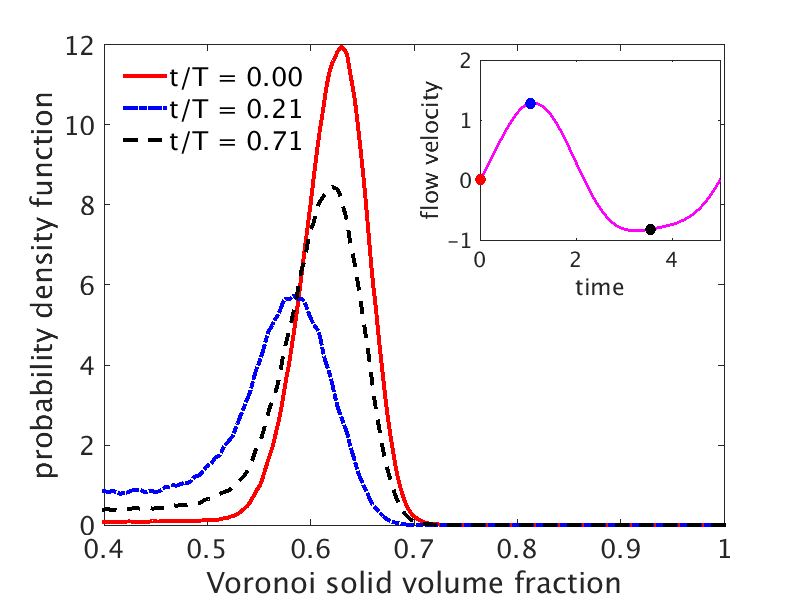}
  }
  \subfloat[ Case 2 (coarse sand), probability density]{
  \includegraphics[width=0.45\textwidth]{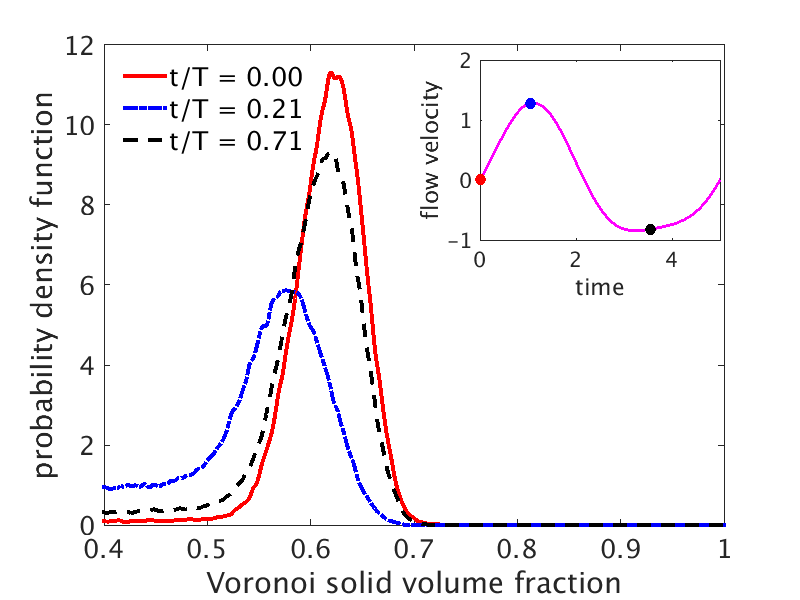}
  }
  \caption{The horizontally-averaged vertical profiles and probability density functions of Voronoi
  volume fraction of Cases 1 and 2 at three representative phases: (1) $t/T = 0$, (2) $t/T = 0.21$,
  and (3) $t/T = 0.71$.}
  \label{fig:voronoii-all}
\end{figure}

The horizontally averaged profiles of Voronoi volume fraction at three representative phases in the
oscillatory sheet-flow are shown in Figures~\ref{fig:voronoii-all}(a) and (b).  For both medium and
coarse sands, the Voronoi volume fractions at the phases corresponding to the maximum
onshore/offshore sediment fluxes ($t/T = 0.21$ and 0.71) are more diffusive than the phases
corresponding to the minimum sediment flux ($t/T = 0.00$).  Here, diffusive means the gradient of
the micromechanic variable is smaller. The Voronoi volume fraction becomes diffusive because the
mixing of solid phase and fluid phase at large fluid velocities. In addition, the sediment particles
occupy more space when suspended in the flow.  The Voronoi volume fraction profile at the maximum
onshore-directed fluid flux ($t/T = 0.21$) is more diffusive than that at the maximum
offshore-directed fluid flux ($t/T = 0.71$). This can be attributed to the fact that the number of
suspended particles is larger when the onshore-directed flow velocity is larger than the
offshore-directed flow velocity.  In addition, the Voronoi volume fraction for coarse sand is less
diffusive than medium sand at the phases corresponding to the maximum onshore/offshore sediment
fluxes. This is because the inertia of coarse particles is larger and they are less likely to become
suspended.  The probability density functions of distribution of Voronoi volume fraction are also
used to demonstrate the validation of micromechanic variables in Figures~\ref{fig:voronoii-all}(c)
and (d).  It can be seen that the peaks of the probability density function for the sediment bed
corresponding to the minimum fluid flux ($t/T = 0.00$) in both cases are at $\varepsilon_s = 0.62$,
which is consistent with the volume fraction for poured random packing~\citep{dullien12pm}.  When
the fluid flux increases, the sediment particles on the bed become suspended and occupy more space.
Hence, more particles have larger Voronoi cells, and fewer particles are closely packed. Therefore,
the peak value in the probability density function moves leftward and decreases in magnitude.

\begin{figure}[htbp]
  \centering
  \subfloat[ Case 1 (medium sand), vertical profile ]{
  \includegraphics[width=0.45\textwidth]{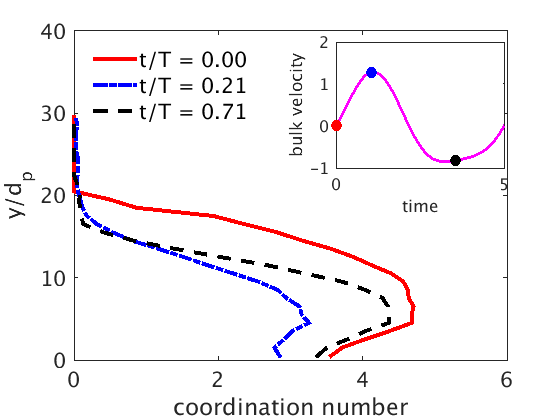}
  }
  \subfloat[ Case 2 (coarse sand), vertical profile ]{
  \includegraphics[width=0.45\textwidth]{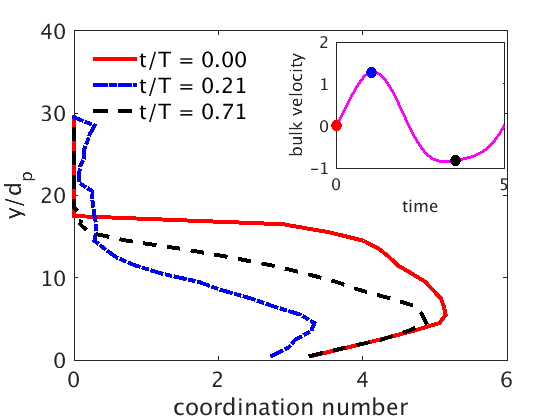}
  }
  \vspace{0.1 in}
  \subfloat[ Case 1 (medium sand), probability density]{
  \includegraphics[width=0.45\textwidth]{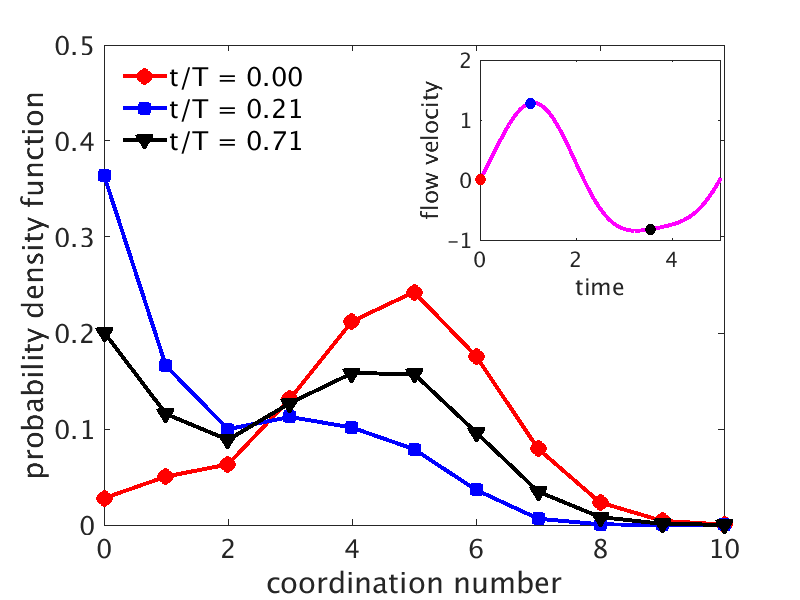}
  }
  \subfloat[ Case 2 (coarse sand), probability density]{
  \includegraphics[width=0.45\textwidth]{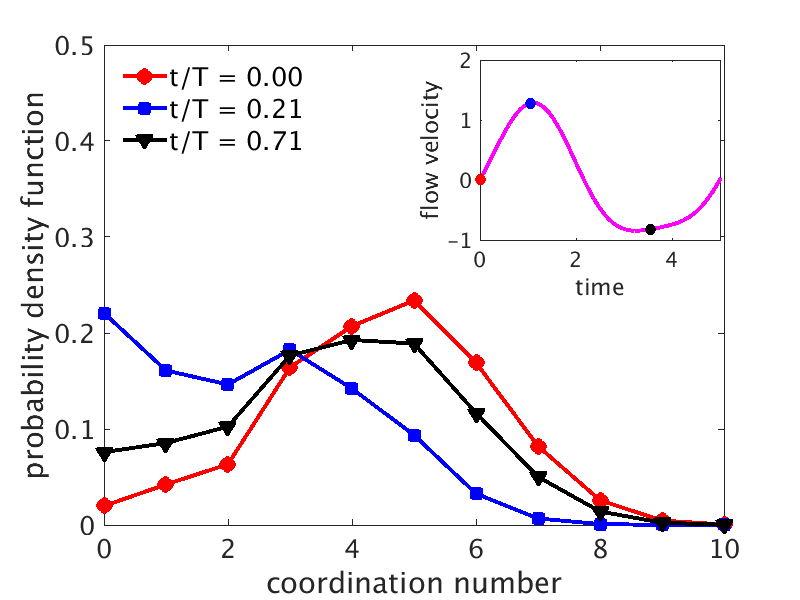}
  }
  \caption{The horizontally-averaged vertical profiles and probability density functions of
  coordination number of Cases 1 and 2 at three representative phases: (1) $t/T = 0$, (2) $t/T =
0.21$, and (3) $t/T = 0.71$.}
  \label{fig:CN-all}
\end{figure}

The coordination number (CN) is defined as the number of nearest neighbors of the sediment particle.
This quantity is useful in the evaluation of the structural properties of the sediment particles,
for example, the packing and permeability of the sediment bed. The horizontally averaged
coordination number profiles for Cases 1 and 2 are shown in Figures~\ref{fig:CN-all}(a) and (b),
respectively. It can be seen that the coordination number decreases at the phases corresponding to
the maximum onshore/offshore sediment fluxes, which is because the suspended particles have less
contact when suspend. Moreover, the probability density functions are shown in
Figures~\ref{fig:CN-all}(c) and (d). The peak of the probability density functions is located at
coordination number CN = 5 at $t/T = 0.00$, which is consistent with results obtained in
the DEM simulations of a rotating drum~\citep{yang23ma}.  Figures~\ref{fig:CN-all}(c) and (d) also
show the variation of the coordination number at different phases in the oscillatory cycle.  Similar
with those of the volume fraction, the peaks of the probability density move leftward when the
sediment transport rate increases.  Moreover, a significant increase for the probability density
function of the coordination number CN~=~0~and~1 can be seen in Figures~\ref{fig:CN-all}(c) and (d),
which is due to the suspension of sediment particles. The increase of the probability density of
coarse sand is less than that for the medium sand, which is due to the fact that coarse particles
are less likely to become suspended.  Compared to the prediction of the coordination number by
CFD--DEM, the two-fluid model uses binary collision assumptions and the coordination number CN = 1
is assumed constant~\citep{cheng14ts}. It can be seen from Figures~\ref{fig:CN-all}(a) and (b) that
the coordination number under the sediment bed is much larger than 1. Therefore, the binary
collision assumption underestimates the contact of the sediment particles. 

\begin{figure}[htbp]
  \centering
  \subfloat[ Case 1 (medium sand)]{
  \includegraphics[width=0.45\textwidth]{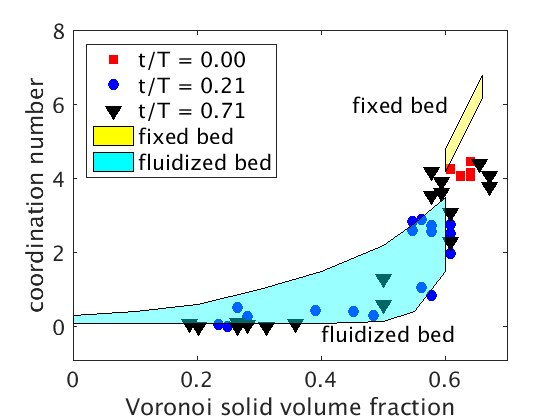}
  }
  \subfloat[ Case 2 (coarse sand)]{
  \includegraphics[width=0.45\textwidth]{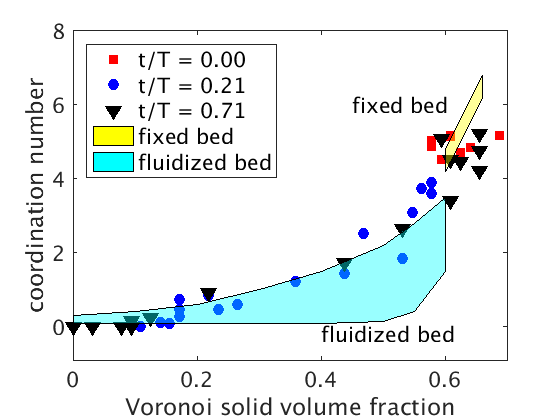}
  }
  \caption{The relationship between the coordination number and Voronoi volume fraction. The shade
    is obtained from the fluidized bed simulations by~\cite{hou12mm}.}
  \label{fig:CN-epsilon}
\end{figure}

The relationship between the Voronoi volume fraction and the coordination number is shown in
Figure~\ref{fig:CN-epsilon}. The relationship obtained in sediment transport is compared to the
results obtained in the CFD--DEM simulations of fluidized bed flow~\citep{hou12mm}. It can be seen
that at the phase corresponding to minimum fluid flux ($t/T$ = 0.00), the correlation of Voronoi
volume fraction and coordination number is consistent with the results obtained in fixed bed regime
in the fluidized bed flow. At the phases corresponding to the maximum onshore/offshore fluid flux
($t/T$ = 0.21 and 0.71), this correlation is consistent with fluidized bed regime. The ratio of
coordination number and Voronoi volume fraction of coarse sand is larger than that of medium sand,
which indicates that the fraction of fluidized medium sand in the seabed is larger than that of
coarse sand. This ratio is even larger when the fluid flow is offshore-directed for coarse sand,
which is described as partially-fluidized by~\cite{hou12mm}. This is because the coarse particles
are less likely to move when the offshore-directed fluid flux is relatively small.

\begin{figure}[htbp]
  \centering
  \subfloat[ Case 1 (medium sand)]{
  \includegraphics[width=0.45\textwidth]{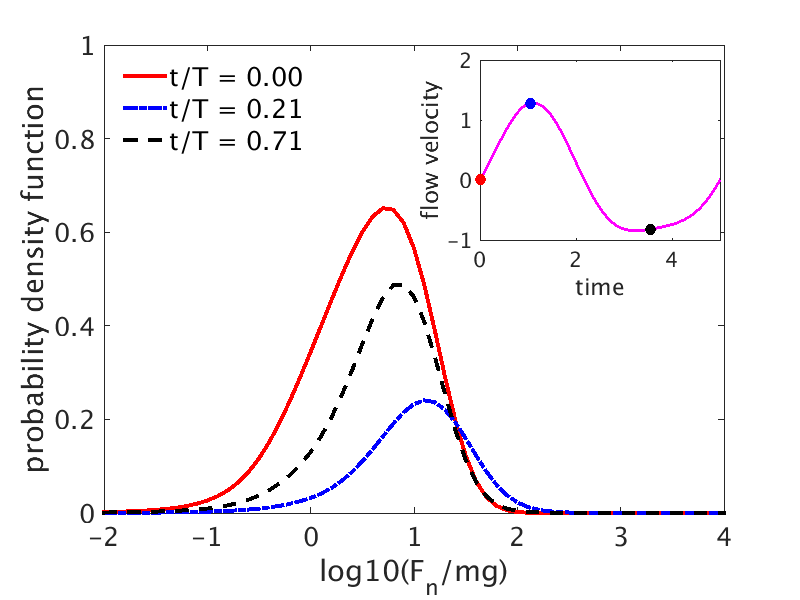}
  }
  \subfloat[ Case 2 (coarse sand)]{
  \includegraphics[width=0.45\textwidth]{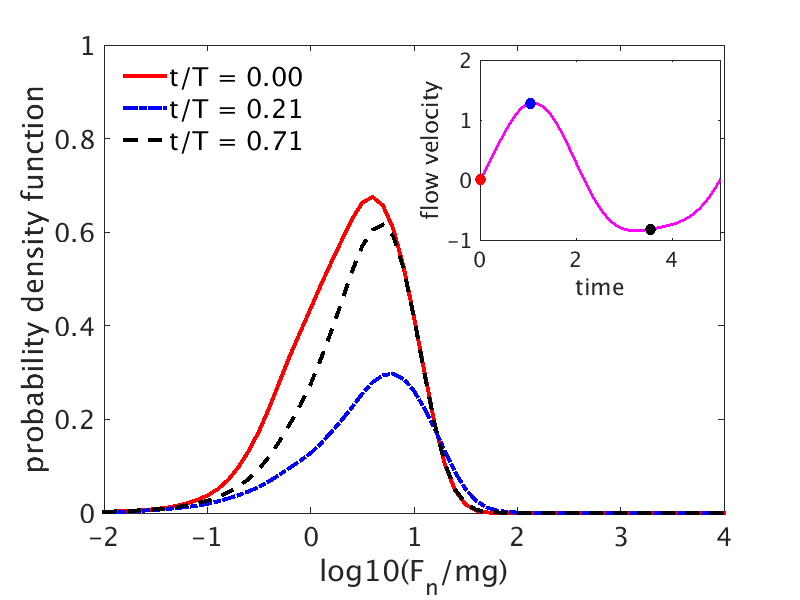}
  }
  \caption{Probability density functions of the contact force at different phases in oscillatory
  flow for Cases 1 and 2.}
  \label{fig:FN-pdf}
\end{figure}

\begin{figure}[htbp]
  \centering
  \subfloat[ Case 1 (medium sand), $t/T = 0$]{
  \includegraphics[width=0.8\textwidth]{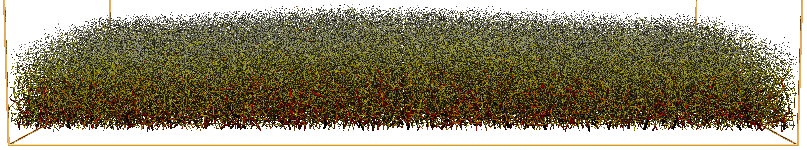}
  }
  \vspace{0.0in}
  \subfloat[ Case 2 (coarse sand), $t/T = 0$ ]{
  \includegraphics[width=0.8\textwidth]{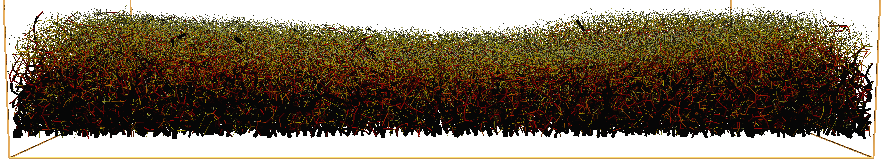}
  }
  \vspace{0.0in}
  \subfloat[ Case 1 (medium sand), $t/T = 0.21$]{
  \includegraphics[width=0.8\textwidth]{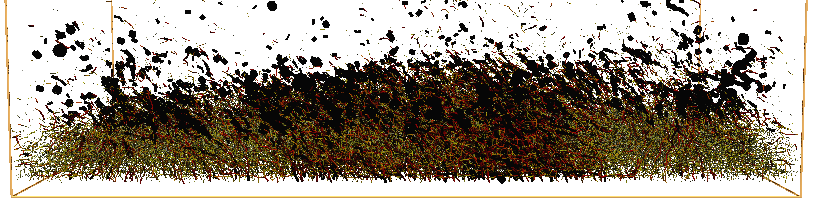}
  }
  \vspace{0.0in}
  \subfloat[ Case 2 (coarse sand), $t/T = 0.21$]{
  \includegraphics[width=0.8\textwidth]{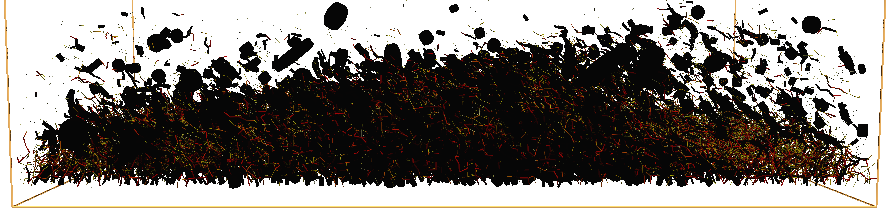}
  }
  \vspace{0.0in}
  \includegraphics[width=0.4\textwidth]{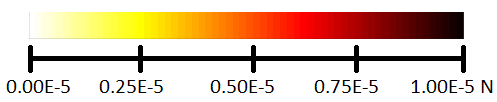}
  \caption{The force network of the sediment transport in oscillatory flow for Cases 1 and 2 at two
  representative phases in the oscillatory flow cycle: (1) $t/T = 0$ (fixed bed with zero sediment
  flux); (2) $t/T = 0.21$ (maximum sediment flux).  The color and the diameter of the cylinder
  denote the magnitude of the force.}
  \label{fig:FN-all}
\end{figure}

Another micromechanic variable is the normal contact force between sediment particles.
Figure~\ref{fig:FN-pdf} shows the probability density functions of the normal contact forces. The
contact force is normalised by the weight of the particle. The peaks of the density functions are
located approximately at $\xmbs{F_n}/m\xmbs{g} = 10$, which is consistent with previous DEM
simulations of a rotating drum~\citep{yang23ma}. These peaks of the probability density decrease in
magnitude when $t/T = 0.21$ and 0.71 because the particles located near the top become suspended.
Since medium sand particles are more likely to become suspended than coarse particles, the decrease
of the magnitude of the peak is larger.  Figure~\ref{fig:FN-pdf} also shows the increase of the
normal contact force when the fluid flux increases. To investigate this increase of the contact
force, the contact force network is visualized to determine the force chain formation in the
sediment bed, which is shown in Figure~\ref{fig:FN-all}.  The contact force network is the plot of
the contact force between contacting particles. The contact force network can illustrate the
propagation of the contact force in granular materials.  In Figure~\ref{fig:FN-all}, the color and
diameter of the cylinder denote the magnitude of the force.  It can be seen in
Figure~\ref{fig:FN-all} that the contact force for fixed sediment bed increases at the bottom due to
the effect of gravity force.  On the other hand, for moving sediment bed, the contact force of some
suspended particles is large. The suspended particles have larger velocity, and the probability to
have large relative velocity between the particles is higher.  Therefore, the contact force of some
suspended particles is significantly larger than particles rolling and sliding on the sediment bed.
It can be also seen in Figures~\ref{fig:FN-all}(a) and (b) that the contact forces between coarse
sand particles are larger than those in medium sand at the phases that the fluid flux is zero.  This
can be explained by the fact that when flow flux is zero, most sediment particles are not moving on
the sediment bed. Since the relative velocity between fluid and sediment particle is small, the drag
force on the sediment particle is small compared to particle weight. In addition, the pressure
gradient force is almost equal to the buoyancy force since the flow is stationary.  The resultant of
the contact forces on each particle balances the particle's submerged weight, which is the sum of
gravity and buoyancy.  Since the submerged weight of coarse sand is larger than that of medium sand,
the inter-particle contact force on coarse sand is larger.

\section{Conclusion}
\label{sec:conclude}

In this work, numerical simulations of sediment transport in oscillatory sheet-flow are performed by
using CFD--DEM solver \textit{SediFoam}. The results obtained by using CFD--DEM simulations are
compared to the experimental data of coarse and medium sands.  The comparisons have demonstrated
that predictions of fluid velocity, the sediment flux, and the sediment transport rate are
consistent with the experimental measurements at different phases in the oscillatory flow cycle. The
consistency between the experimental data and the results from the numerical simulation demonstrate
that CFD--DEM is capable of simulating the sediment transport problems in oscillatory flow for
coarse and medium sands. 

In addition, the micromechanics of sediment bed are studied, including the Voronoi volume fraction,
the coordination number, and the particle contact force. It is demonstrated that the behavior of the
micromechanic variables varies significantly at different phases in the oscillatory flow, which
indicates the microscopic structure of sediment bed is dependent on the fluid flux. Moreover, the
micromechanic variables of the sediment bed are similar to the fluidized bed at high sediment
transport rate; whereas the variables are similar to the fixed bed at low sediment transport rate.
From the prediction of the CFD--DEM model, we observed that the coordination number in rapid sheet
flow layer is larger than one, which indicates that a typical particle in the sediment layer is in
contact with more than one particles, and thus the binary collision model commonly used in two-fluid
model may underestimate the contact between the particles.

\section{Acknowledgment}

The computational resources used for this project were provided by the Advanced Research Computing
(ARC) of Virginia Tech, which is gratefully acknowledged. The authors would like to thank for
Dr.~Gupta for the code and technical assistance in the post-processing of the granular
micromechanics.  The authors gratefully acknowledge partial funding of graduate research
assistantship from the Institute for Critical Technology and Applied Science (ICTAS, Grant number
175258).

\appendix
\setcounter{secnumdepth}{0}
\section{Appendix. A}

\renewcommand\thefigure{A.\arabic{figure}}
\renewcommand\theequation{A.\arabic{equation}}
\setcounter{figure}{0}

The definitions and the post-processing procedures of the micromechanic variables, including Voronoi
solid volume fraction and coordination number, are detailed in this appendix. 

\subsection{A.1 Voronoi Volume Fraction}

\begin{figure}[htbp]
  \centering
  \includegraphics[width=0.35\textwidth]{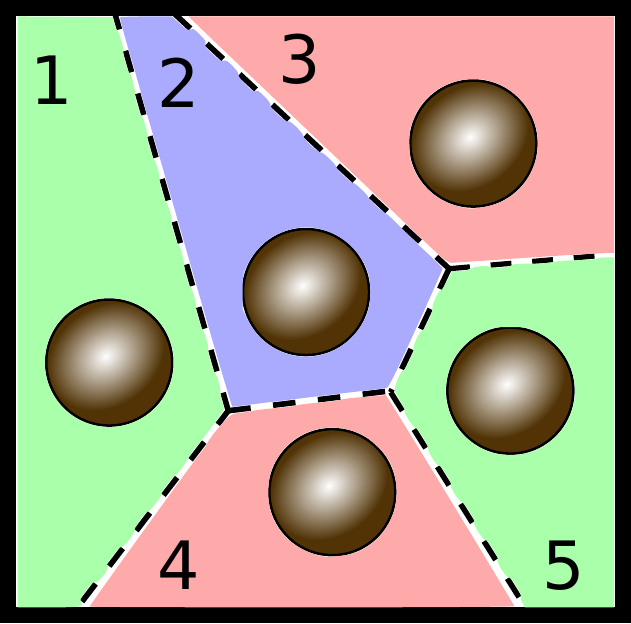}
  \caption{2D Voronoi cell of sediment particles.}
  \label{fig:Voronoi-layout}
\end{figure}

The Voronoi cell is defined according to the Voronoi diagram in Figure~\ref{fig:Voronoi-layout}. The
Voronoi cell of a spherical sediment particle is the region that is closer to the center of this
spherical particle than other points in the system~\citep{yang02vt}.  It can be seen in the figure
that the regions highlighted with filled color are the Voronoi cells, numbering from 1 to 5. The 3D
Voronoi cells are created by taking pairs of points that are close together and drawing a plane that
is equidistant between them and perpendicular to the line connecting them.  This region is
considered a free volume for the particle to move. The Voronoi solid volume fraction is the ratio
between the volume of the particle and the Voronoi cell. The calculation of the Voronoi solid volume
fraction is based on the Voronoi package of LAMMPS~\citep{lammps}.

\subsection{A.2 Coordination Number}
\begin{figure}[htbp]
  \centering
  \includegraphics[width=0.6\textwidth]{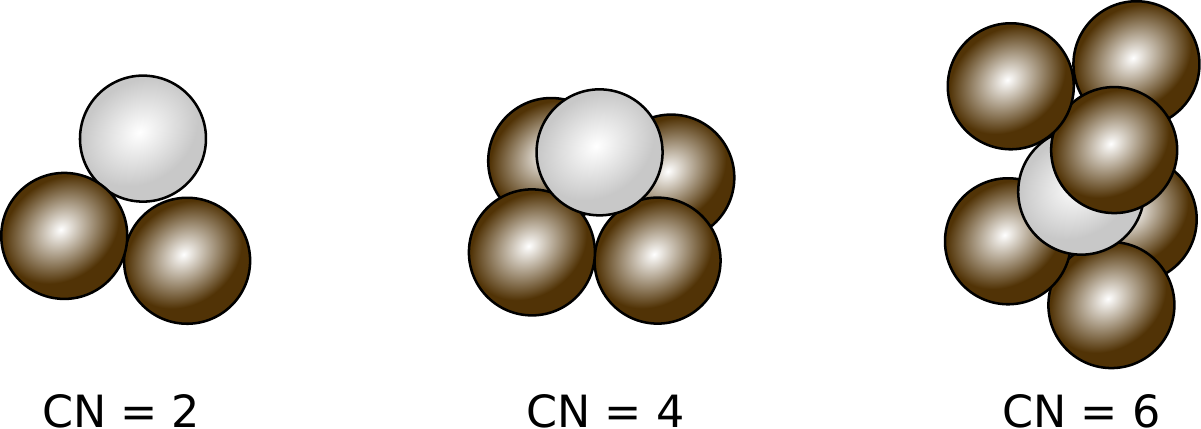}
  \caption{The coordination number (CN) of the light particle in different packing regimes.}
  \label{fig:CN-layout}
\end{figure}

The coordination number of a sediment particle is defined as the number of nearest neighbors of the
particle.  Figure~\ref{fig:CN-layout} shows the coordination number of different
packing configurations. The coordination number is useful in the investigation of the momentum and
heat transfer of the particles. This quantity is also useful in particulate flow in the evaluation
of the structural properties of packing particles, such as tensile strength, thermal conductivity,
solid-solid reactions, phase formation, and permeability~\citep{yi11cn}.

\subsection{A.3 Spatial Averaging Procedure}
The vertical profiles of the Voronoi volume fraction and coordination number are obtained by
averaging the information of individual particles. According to~\cite{gupta15vv}, the quantities
are averaged by using:
\begin{equation}
  \zeta(\mathbf{x}) = \sum_{i=1}^{N_p} \mathcal{Z}_i h_i (\mathbf{x_i}),
 \label{eq:superpo}
\end{equation}
where $\zeta$ is the quantity obtained after the averaging procedure; $\mathcal{Z}_i$ is the same
quantity of the sediment particle before the averaging; $N_p$ is the number of particles in the
cell; $\mathbf{x}$ is the location of the cell; $\mathbf{x_i}$ is the location of particle $i$. The
function $h_i (\mathbf{x_i})$ is defined as:
\begin{equation}
   h_i(\mathbf{x_i}) = \frac{1}{(b^2 {\pi})^{3/2}} \exp \left[ - \frac{(\mathbf{x}-\mathbf{x}_i)^T
   (\mathbf{x}-\mathbf{x}_i) }{b^2} \right],
    \label{eq:hi}
\end{equation}
where $b = 4d_p$ is the bandwidth of the averaging.  In this paper, the resolution of the
post-processing grid is $1\times30\times1$ (in streamwise, vertical, and lateral directions,
respectively). The size of the cell in the vertical direction is the particle diameter $d_p$.
Therefore, the maximum height of the post-processing grid is $30d_p$, which is higher than the
maximum height reached by the sediment particles. Since the resolution of post-processing mesh in
streamwise and lateral direction is 1, the results are horizontally-averaged.

\end{document}